\documentclass[showpacs,amsmath,pre,floatfix]{revtex4}
\usepackage{amsmath,amsfonts,amssymb}
\usepackage{graphicx}
\usepackage{pspicture}

\begin{document}
\title{Metastable states and $T=0$ hysteresis in the random-field
Ising model on random graphs.}

\author{F. Detcheverry, M. L. Rosinberg, and G. Tarjus}
\affiliation{Laboratoire de Physique  Th\'eorique des Liquides,  Universit\'e Pierre et
Marie Curie, 4 place Jussieu, 75252 Paris Cedex 05, France}

\begin{abstract}
We study the ferromagnetic  random-field Ising model on  random 
graphs of fixed connectivity $z$ (Bethe lattice) in the presence of an
external magnetic field $H$. We compute the number of single-spin-flip
stable configurations with a given   magnetization $m$ and study   the
connection between  the distribution of these metastable states in the $H-m$ plane
(focusing on  the region where the number  is exponentially large) and
the shape  of the saturation hysteresis loop obtained by cycling the
field  between $-\infty$ and $+\infty$ at $T=0$.  The  annealed complexity $\Sigma_A(m,H)$ is
calculated  for $z=2,3,4$ and the  quenched  complexity $\Sigma_Q(m,H)$ for
$z=2$.  We prove  explicitly     for $z=2$ that the contour
$\Sigma_Q(m,H)=0$ coincides with the saturation loop. On the other hand, we  show  that 
$\Sigma_A(m,H)$  is irrelevant for describing, even qualitatively, the observable hysteresis properties of the system.

\end{abstract}
\maketitle

\def\be{\begin{equation}}
\def\ee{\end{equation}}
\def\bea{\begin{align}}
\def\eea{\end{align}}
\def\ss{\scriptscriptstyle}
\section{Introduction}

The dynamical properties of random and frustrated systems are dominated
at low temperature by the presence of a
multitude of metastable states.
 This complex energy
landscape is responsible for the
strong hysteresis effects observed in a variety of physical systems
(magnets, superconductors, fluids in porous media,\ldots) under the action of an external driving field.
Much insight into such behavior has been gained
recently through studying the zero-temperature random-field Ising model (RFIM) 
in an external magnetic field, which is a simple prototype of a disordered
system far from equilibrium. The energy landscape of the model
may be explored by changing the field adiabatically, which
results in a series of discrete Barkhausen jumps (avalanches) in the
magnetization, each jump corresponding to a sudden move to a new
metastable configuration. A remarkable feature of the  RFIM at $T=0$
is the existence of a disorder-induced phase transition in the
magnetization hysteresis loop obtained
by cycling the field from one saturated state to the oppositely
magnetized state and back. In three and higher
dimensions, there is  a transition between a continuous loop at
strong disorder and a discontinuous loop with  a
 macroscopic jump in
the magnetization  at low disorder\cite{S1993}. The jump
(corresponding to a macroscopic avalanche in the system) disappears at some
critical value of the disorder where universal scaling behavior is
observed. This nonequilibrium phase
transition has been studied in detail by renormalization
group and by numerical simulations on hypercubic lattices \cite{DS1996}. Exact
analytical calculations have also been performed in one dimension\cite{S1996}
 and generalized  to a Bethe
lattice\cite{DSS1997}: in this case, the transition only exists for
$z\geq4$, where $z$ is the connectivity of the lattice. Such
a transition has been observed in thin Co/CoO films\cite{BIJPB2000} and
Cu-Al-Mn alloys\cite{M2003}, and we have recently suggested\cite{DKRT2003} that it may also be associated to the 
change in the adsorption behavior of $^4$He in dilute silica aerogels\cite{TYC1999}.  In all these systems, activated processses are 
too slow to allow thermal equilibration on experimental time scales, which makes the $T=0$ RFIM a relevant model for understanding the  out-of-equilibrium behavior.

There is an expected connection between the shape of the $T=0$
hysteresis loop and the distribution of the metastable states in the
field-magnetization plane\cite{note00}. This issue, however, has only attracted little attention so far\cite{PZZ1999,L2000}. For the RFIM with ferromagnetic interactions, the probability to find metastable states outside the hysteresis loop vanishes in the thermodynamic limit, as shown in Appendix A. But what is the situation  {\it inside} the loop ? Is the entropy density (associated with the number of metastable states) positive everywhere ? The situation is especially intriguing in the low-disorder regime where a region around the origin becomes inaccessible to any field history that starts from the saturated states\cite{CGZ2002}.
The aim of this work is thus to calculate
the number of  metastable configurations  with a fixed magnetization $m$ per spin
as a function of the external field $H$. We consider a Gaussian distribution of the random fields whose width $R$ measures the "strength" of the disorder, and, in order to analyze the problem analytically (at least partially), we work on a Bethe lattice defined here as a random graph of fixed connectivity $z$ \cite{MP2001}, i.e. a random collection of $N$ vertices, each vertex being connected to $z$ other vertices. Such a graph has the structure of a branching tree locally and small loops are rare, with typical size of order $\log N$. Spin models
on random graphs (both with fixed and
fluctuating connectivity) have attracted much attention in recent years because
they can be solved exactly though each spin interacts with a finite
number of  other spins like in real short-range, finite-dimensional
systems. The metastable configurations of the Ising
ferromagnet\cite{LD2001,BS2001,PPR2002} and of spin
glass models\cite{D2000,PPR2002} have been studied in this framework, but, to the best of our knowledge, the ferromagnetic RFIM has not yet been considered (except for $z=2$ and in zero external field\cite{MJWS1993}).  Since there is no plaquette frustration in this case (contrary to spin glasses), loops should not play any significant role in the large-$N$ limit and we expect the number of metastable states on a typical random graph with connectivity $z$ to be the same as in the interior of a Cayley tree with branching ratio $z-1$\cite{note1}.

As is usual with disordered systems, one has to distinguish between 
annealed and quenched averages, the disorder average being here over both the random-graph and the random-field distributions. The quenched average is the appropriate quantity describing the behavior of the typical disorder realizations.
 Specifically, we want to compute both
the {\it average} number $\overline{{\cal N}(m,H)}$ and the {\it typical} number
$e^{\overline{\ln {\cal N}(m,H)}}$ of single-spin-flip stable states with a given magnetization $m$. Since these numbers are expected to
scale exponentially with the number $N$  of spins (at least in some
region of the $H-m$ plane), the basic quantities are the associated entropy densities, i.e. the annealed complexity 
$\Sigma_A(m,H)=\lim_{N \to \infty}1/N \ \ln \overline{{\cal N}(m,H)}$ and the
quenched complexity  $\Sigma_Q(m,H)=\lim_{N \to \infty}1/N \ \overline{\ln {\cal N}(m,H)}$.
 In the following, we present the results for the annealed complexity on  random graphs with connectivity
$z=2,3$ and $4$. We shall see that the qualitative behavior changes for
$z\geq3$, although 
the recursion equations for the RFIM on a Bethe lattice
predict that the macroscopic jump in the hysteresis loop only occurs for $z\geq
4$\cite{DSS1997}. This illustrates that the annealed quantity is unable to describe, even qualitatively, the typical behavior of the system: indeed, the
probabilistic calculation of Ref.\cite{DSS1997} describes the 
behavior of the magnetization which is a self-averaging quantity that 
converges to its typical value when $N \to \infty$. (Note that the crucial distinction between annealed and quenched averages was neglected in Ref.\cite{L2000}.)
The analytical computation of the quenched complexity requires the introduction of replicas and is thus more complicated (alternatively, one could also use the cavity method as in
Ref.\cite{PPR2002}). For this reason, we carry out the complete  calculation of $\Sigma_Q(m,H)$ 
for $z=2$  only. We then  show explicitly that the contour $\Sigma_Q(m,H)=0$ identifies
with the saturation loop of the one-dimensional RFIM  previously calculated  by a
probabilistic method\cite{S1996}: the typical number of metastable states thus ceases to be exponential with the system size exactly on the saturation hysteresis loop. To our knowledge, this is the first time that such a result is proven exactly. 

The rest of the paper is organized as follows. In Sec. II we introduce the model
and describe the averaging procedure over the random graphs.
In Sec. III we present the general formalism for computing the annealed and quenched complexities on such graphs. The analytical and numerical results for the one-dimensional chain ($z=2$) are discussed in Sec. IV. In Sec. V we present the results for the annealed complexity for $z=3$ and $z=4$. We conclude in Sec. VI by discussing the possible behavior of the quenched complexity for $z\ge 4$ in the low-disorder regime. The absence of metastable states outside the saturation hysteresis loop is proven in Appendix A. Appendices B and C are devoted to technical details of the quenched calculation for $z=2$.

\section{Model}

We study the ferromagnetic RFIM described by the Hamiltonian

\be
H=-J\sum_{i<j}n_{ij}s_is_j -\sum_i(H+h_i)s_i \ ,
\ee
where  $J>0$,  $s_i$ ($i=1,\ldots,N$) are Ising spins placed on the vertices of
a random graph, and the  random  fields $h_i$ are drawn  from the
Gaussian probability distribution  $P(h_i)=\exp(-h_i^2/2R^2)/\sqrt{2\pi}R $. The variable $n_{ij}$ is equal to $1$ if the sites  $i,j$ are connected and to $0$ otherwise. 

Following Ref.\cite{D2000}, the  random graphs are built as follows. First, the 
 elements  $n_{ij}$ of the  connectivity  matrix (with $n_{ij}=n_{ji}$, $i<j$) are chosen at random 
from  the probability distribution ${\cal P}(n_{ij})=(1-\gamma/N)\delta(n_{ij};0)+(\gamma/N) \delta(n_{ij};1)$
where $\gamma$ is some arbitrary number of order one and $\delta(p;q)$ is the Kronecker-$\delta$: this defines  an ensemble of
graphs where the local connectivity is a Poisson variable in the large-$N$ limit, like in the Viana-Bray spin glass model\cite{VB1985}. One then selects the subset of graphs where the connectivity of each
site is equal to $z$. The average value of a dynamical variable $A$ is thus given by
\be
\overline{A}=\frac{1}{{\cal M}(N,z,\gamma)}\langle\langle A \prod_{i=1}^{N} \delta(\sum_{j\neq i}n_{ij};z)\rangle_{\gamma}\rangle_h \ ,
\ee
where $\langle .\rangle_{\gamma}$ and $\langle.\rangle_h$ denote the averages with respect
to the distributions ${\cal P}(n_{ij})$ and $P(h_i)$, respectively.  ${\cal M}(N,z,\gamma)$ is the average number of
graphs with connectivity $z$ that are generated for a given $\gamma$. This
quantity was calculated in Ref.\cite{D2000} in the large-$N$ limit:
\be
\lim_{N \to \infty}\frac{1}{N}\ln {\cal M}(N,z,\gamma)=\frac{z}{2}(\ln z+\ln \gamma -1)-\ln z! -\frac{\gamma}{2} \ .
\ee
By construction, the value of $\overline{A}$ is independent of the choice of $\gamma$ in the thermodynamic limit $N \rightarrow \infty$. In the following, we drop the explicit dependence on $z$ and $\gamma$.

\section{Formalism}

In this section we present the general formalism for computing the number of metastable states of the Gaussian RFIM on random graphs. Since the quenched calculation is achieved by introducing replicas and using the identity $\ln x=\lim_{n\to 0}(x^n-1)/n$ , we compute directly the average  $\overline{{\cal N}(m,H)^n}$ for  a general integer $n$, and then treat separately the annealed ($n=1$) and  quenched ($n\rightarrow 0$) cases.  Our method is inspired from Ref.\cite{BS2001} where a global order-parameter function is introduced that allows to adress the quenched calculation.

\subsection{Basic equations}

A configuration is said to be metastable when its energy cannot be
decreased by flipping a single spin, a definition consistent with the  $T=0$
Glauber dynamics used in the previous studies of the nonequilibrium
RFIM\cite{S1993,DS1996,S1996,DSS1997}. Such a configuration is a local minimum of the energy surface where each spin $i$ is aligned with its local field
$f_i$ defined by
\be
f_i=J \sum_j n_{ij} s_j+h_i+H \ .
\ee
The $n$th moment of the number of metastable states averaged over disorder is thus given by
\be
\overline{{\cal N}(H)^n}=\frac{{\cal D}_n(N,H)}{{\cal M}(N)}
\ee
with
\begin{align}
{\cal D}_n(N,H)&= \langle\langle\big[\sum_{s_i=\pm 1}\prod_{i=1}^{N} \Theta(f_is_i)\big]^n\delta(\sum_{j\neq i}n_{ij};z)\rangle_{\gamma}\rangle_h\nonumber\\
&=\langle\langle\mbox{Tr }\prod_{i,a}\Big[\Theta(f_i^as_i^a)\Big]\delta(\sum_{j\neq i}n_{ij};z)\rangle_{\gamma}\rangle_h
\end{align}
where Tr  stands for the summation over all possible values of the $n$-replicated spins $s_i^a$ ($a=1,...,n$), $f_i^a=J \sum_j n_{ij} s_j^a+h_i+H$, and $\Theta(x)$ is the Heaviside step function (i.e. $\Theta(x)=1$ if $x> 0$ and $0$ if $x<0$). Since we consider a continuous distribution of the random fields, there is no need to specify the value $\Theta(0)$: the local field $f_i$ is almost surely different from zero and the system has no marginally stable states whose energy does not change under any single-spin flip.  We then use integral representations of the Dirac-$\delta$ and Kronecker-$\delta$
to rewrite Eq.(6) as
\begin{widetext}
\begin{align}
&  {\cal D}_n(N,H)=\langle\langle \mbox{Tr } \int \prod_{i,a} \Big[dx_i^a  \delta(x_i^a-f_i^a)\Theta(x_i^as_i^a)\Big] \delta(\sum_{j\neq i}n_{ij};z)\rangle_{\gamma}\rangle_h \nonumber \\
&= \langle\langle\mbox{Tr }
\int \prod_{i,a} \Big[dx_i^ady_i^a\Theta(x_i^as_i^a)\Big] e^{\sum_i{\bf y}_i.({\bf x}_i-{\bf f}_i)}\delta(\sum_{j\neq i}n_{ij};z)\rangle_{\gamma}\rangle_h
\nonumber \\
&= \langle\langle\mbox{Tr }
\int \prod_{i,a} \Big[dx_i^ady_i^ad\lambda_i\Theta(x_i^as_i^a)\Big] e^{\sum_{i}[{\bf y}_i.({\bf x}_i-{\bf f}_i)-\lambda_i(\sum_{j\neq i}n_{ij}-z)]}\rangle_{\gamma}\rangle_h
\end{align}
\end{widetext}
where the bold symbols denote vectors in replica space, e.g. ${\bf x}_i=(x_i^1,x_i^2,...,x_i^n)$. The integration ranges are $[-\infty,+\infty]$  for $x_i^a$, $[-i\infty,+i\infty]$ for $y_i^a$,
$ [0,2i\pi]$ for $\lambda_i$, and for brevity the normalisation factors $1/(2i\pi)$ are absorbed in the infinitesimal elements $dy_i^a$ and $d\lambda_i$.  Inserting the definition of $f_i^a$ (with ${\bf h}_i=h_i{\bf 1}, {\bf H}=H{\bf 1}$) and performing the average over the distribution ${\cal P}(n_{ij})$ yields
\begin{align}
{\cal D}_n(N,H)&= \langle\mbox{Tr }
\int \prod_{i,a} \Big[dx_i^ady_i^ad\lambda_i\Theta(x_i^as_i^a)\Big]
e^{\sum_i[\lambda_iz+{\bf y}_i.({\bf x}_i-{\bf h}_i-{\bf H})]}\nonumber\\
&\times\prod_{i<j}(1-\frac{\gamma}{N}+\frac{\gamma}{N}e^{-\lambda_i-\lambda_j-J({\bf y}_i.{\bf s}_j+{\bf y}_j.{\bf s}_i)})\rangle_h
\nonumber \\
&=\langle\mbox{Tr}\int\prod_{i,a} \Big[ dx_i^ady_i^ad\lambda_i\Theta(x_i^as_i^a)\Big]
e^{\sum_i[\lambda_iz+{\bf y}_i. ({\bf x}_i-{\bf h}_i-{\bf H})]} \nonumber\\
&\times\exp(-\frac{\gamma N}{2}+\frac{\gamma}{2N}\sum_{i,j}e^{-\lambda_i-\lambda_j-J({\bf y}_i.{\bf s}_j+{\bf y}_j.{\bf s}_i)})\rangle_h
\end{align}
where terms that are negligible in the thermodynamic limit have been dropped in the last exponential.

Like in  Ref.\cite{BS2001} we now introduce an "order-parameter" function $c(\mbox{\boldmath $\sigma$},\mbox{\boldmath $\tau$})=(1/N)\sum_i\delta(\mbox{\boldmath$s$}_i;\mbox{\boldmath $\sigma$}) \exp[-(\lambda_i+J\mbox{\boldmath $y$}_i.\mbox{\boldmath $\tau$})]$, where  $\mbox{\boldmath $\sigma$}$ and $\mbox{\boldmath $\tau$}$ are binary vectors in replica space (i.e. $\sigma^a,\tau^a=\pm1$).
Then,
\be
\exp\big[-\frac{\gamma N}{2}+\frac{\gamma}{2N}\sum_{i,j}e^{-\lambda_i-\lambda_j-J({\bf y}_i.{\bf s}_j+{\bf y}_j.{\bf s}_i)}\big]=\exp[-\frac{\gamma N}{2}+\frac{\gamma N}{2}\sum_{\mbox{\boldmath $\sigma$},\mbox{\boldmath $\tau$}}c(\mbox{\boldmath $\sigma$},\mbox{\boldmath $\tau$})c(\mbox{\boldmath $\tau$},\mbox{\boldmath $\sigma$})] \ ,
\ee
a relation that we use to decouple the sum over the site labels $i$ and $j$ and perform the trace over a single spin only in each replica. As outlined in the Appendix of Ref.\cite{BS2001}, the $2n$-variable function $c(\mbox{\boldmath $\sigma$},\mbox{\boldmath $\tau$})$ may be inserted in Eq. (8) via integrals over $\delta$-functions represented by Fourier integrals over other auxiliary functions. After some formal manipulations, we obtain
\begin{align}
&  {\cal D}_n(N,H)=\int \prod_{\mbox{\boldmath $\sigma$},\mbox{\boldmath $\tau$}}\big[ dc(\mbox{\boldmath $\sigma$},\mbox{\boldmath $\tau$})\big] \exp\{-\frac{\gamma N}{2}\sum_{\mbox{\boldmath $\sigma$},\mbox{\boldmath $\tau$}}c(\mbox{\boldmath $\sigma$},\mbox{\boldmath $\tau$})c(\mbox{\boldmath $\tau$},\mbox{\boldmath $\sigma$}) -\frac{\gamma N}{2}\}
\nonumber \\
&\times \Big[\sum_{\mbox{\boldmath $\sigma$}}\int P(h) dh\int d{\bf x} d{\bf y}d\lambda \exp\{\lambda z+{\bf y}.({\bf x}-{\bf h}-{\bf H})+\gamma e^{-\lambda}\sum_{\mbox{\boldmath $\tau$}} c(\mbox{\boldmath $\tau$},\mbox{\boldmath $\sigma$})e^{-J{\bf y}.\mbox{\boldmath $\tau$}}\}\prod_a\Theta(x^a \sigma^a)\Big]^N
\end{align}
which may be computed in the thermodynamic limit as a saddle-point integral. This leads to
\be
\lim_{N \to \infty}\frac{1}{N} \ln {\cal D}_n(N,H)=\max_{\mbox{c(\boldmath $\sigma$},\mbox{\boldmath $\tau$})} \{-\frac{\gamma}{2}\sum_{\mbox{\boldmath $\sigma$},\mbox{\boldmath $\tau$}}c(\mbox{\boldmath$\sigma$},\mbox{\boldmath$\tau$})c(\mbox{\boldmath$\tau$},\mbox{\boldmath$\sigma$}) -\frac{\gamma}{2}+\Lambda[c]\} \ ,
\ee
with
\be
\Lambda[c]=\ln \sum_{\mbox{\boldmath $\sigma$}}\int P(h) dh\int d{\bf x} d{\bf y}d\lambda\exp\{\lambda z+{\bf y}.({\bf x}-{\bf h}-{\bf H})+\gamma e^{-\lambda}\sum_{\mbox{\boldmath $\tau$}} c(\mbox{\boldmath $\tau$},\mbox{\boldmath $\sigma$})e^{-J{\bf y}.\mbox{\boldmath $\tau$}}\}\prod_a\Theta(x^a \sigma^a) \ .
\ee
Using the identity\cite{LD2001}
\be
\frac{1}{2i\pi}\int_0^{2i\pi} d\lambda e^{\lambda z+\alpha e^{-\lambda}}=\frac{\alpha^z}{z!} \ ,
\ee
we get
\begin{align}
\Lambda[c]&=\ln\frac{\gamma^z}{z!}+\ln \sum_{\mbox{\boldmath $\sigma$}}\int P(h) dh\int d {\bf x} d {\bf y} \ e^{{\bf y}.({\bf x}-{\bf H}-{\bf h})}\nonumber\\
&\times \sum_{\mbox{\boldmath $\tau$}_1...\mbox{\boldmath $\tau$}_z}c(\mbox{\boldmath $\tau$}_1,\mbox{\boldmath $\sigma$})...c(\mbox{\boldmath $\tau$}_z,\mbox{\boldmath $\sigma$})e^{-J{\bf y}.[\mbox{\boldmath $\tau$}_1+\mbox{\boldmath $\tau$}_2+...\mbox{\boldmath $\tau$}_z]}\prod_a\Theta(x^a \sigma^a)\ ,
\end{align}
and the integrations over  $\bf y$ (giving $\delta$ functions again) and then over $\bf x$ may be performed to finally obtain
\be
\Lambda[c]=\ln\frac{\gamma^z}{z!}+\ln \sum_{\mbox{\boldmath $\sigma$},\mbox{\boldmath $\tau$}_1...\mbox{\boldmath $\tau$}_z}c(\mbox{\boldmath $\tau$}_1,\mbox{\boldmath $\sigma$})...c(\mbox{\boldmath $\tau$}_z,\mbox{\boldmath $\sigma$})\int P(h)dh \prod_a\Theta[(H+h+J\sum_{l=1}^z\tau^a_l)\sigma^a] \ .
\ee 
Note that  the random field distribution $P(h)$  has not yet been specified at this stage.

It is straightforward to generalize the preceding calculation for a fixed magnetization $m$. ${\cal D}_n(N,m,H)$ is now defined as
\be
{\cal D}_n(N,m,H)= \langle\langle\Big[\sum_{s_i=\pm 1}\prod_{i=1}^{N} \Theta(h_i^{loc}s_i)\delta(\sum_i s_i-Nm)\Big]^n\delta(\sum_{j\neq i}n_{ij}-z)\rangle_{\gamma}\rangle_h
\ee
and the $\delta$-function constraint on $\sum_is_i$ may be relaxed by introducing a Lagrange multiplier $g$ coupled to the magnetization (see, e.g, Ref.\cite{LD2001}). The whole calculation is quite similar to the one presented above and Eq.(11) becomes
\be
\lim_{N \to \infty}\frac{1}{N} \ln {\cal D}_n(N,m,H)=\max_{\mbox{c(\boldmath $\sigma$},\mbox{\boldmath $\tau$)},\mbox{\boldmath $g$}} \{-\frac{\gamma}{2}\sum_{\mbox{\boldmath $\sigma$},\mbox{\boldmath $\tau$}}c(\mbox{\boldmath$\sigma$},\mbox{\boldmath$\tau$})c(\mbox{\boldmath$\tau$},\mbox{\boldmath$\sigma$}) -\frac{\gamma}{2}-m\sum_ag^a+\Lambda[c,{\bf g}]\} \ ,
\ee
with
\be
\Lambda[c,{\bf g}]=\ln\frac{\gamma^z}{z!}+\ln \sum_{\mbox{\boldmath $\sigma$},\mbox{\boldmath $\tau$}_1...\mbox{\boldmath $\tau$}_z}c(\mbox{\boldmath $\tau$}_1,\mbox{\boldmath $\sigma$})...c(\mbox{\boldmath $\tau$}_z,\mbox{\boldmath $\sigma$})\int P(h)dh \prod_a e^{g^a \sigma^a}\Theta[(H+h+J\sum_{l=1}^z\tau^a_l)\sigma^a] 
\ee
where ${\bf g}=(g^1,g^2,...,g^n)$. As can be seen immediately, Eqs. (11) and (15) are recovered when ${\bf g}={\bf 0}$ and thus Eqs. (17-18) may be used as the starting point of further analysis.

Extremization  with respect to $c(\mbox{\boldmath $\sigma$},\mbox{\boldmath $\tau$})$ readily gives
\be
c(\mbox{\boldmath $\sigma$},\mbox{\boldmath $\tau$})= e^{-\Lambda}\frac{\gamma^{z-1}}{(z-1)!}\sum_{\mbox{\boldmath $\tau$}_1...\mbox{\boldmath $\tau$}_{z-1}}c(\mbox{\boldmath $\tau$}_1,\mbox{\boldmath $\sigma$})...c(\mbox{\boldmath $\tau$}_{z-1},\mbox{\boldmath $\sigma$})F_z(\mbox{\boldmath $\sigma$},\mbox{\boldmath $\tau$},\mbox{\boldmath $\tau$}_1...\mbox{\boldmath $\tau$}_{z-1})
\ee
with
\be
F_z(\mbox{\boldmath $\sigma$},\mbox{\boldmath $\tau$},\mbox{\boldmath $\tau$}_1...\mbox{\boldmath $\tau$}_{z-1})=\int P(h)dh \prod_a  e^{g^a \sigma^a}\Theta[(H+h+J(\tau^a+\sum_{l=1}^{z-1}\tau^a_l))\sigma^a] \ ,
\ee
which in turn yields the normalization condition
\be
\frac{\gamma}{z}\sum_{\mbox{\boldmath $\sigma$},\mbox{\boldmath $\tau$}}c(\mbox{\boldmath $\sigma$},\mbox{\boldmath $\tau$})c(\mbox{\boldmath $\tau$},\mbox{\boldmath $\sigma$})=1 \ .
\ee
Likewise, extremization with respect to  $g^a$ yields
\begin{align}
m&=\partial\Lambda[c,{\bf g}]/ \partial g^a\nonumber\\
&=e^{-\Lambda}\frac{\gamma^z}{z!}\sum_{\mbox{\boldmath $\sigma$},\mbox{\boldmath $\tau$}_1...\mbox{\boldmath $\tau$}_z}\sigma^ac(\mbox{\boldmath $\tau$}_1,\mbox{\boldmath $\sigma$})...c(\mbox{\boldmath $\tau$}_z,\mbox{\boldmath $\sigma$})\int P(h)dh \prod_a e^{g^a \sigma^a}\Theta[(H+h+J\sum_{l=1}^z\tau^a_l)\sigma^a] \nonumber\\
&=\sum_{\mbox{\boldmath $\sigma$},\mbox{\boldmath $\tau$}}
\sigma^a c(\mbox{\boldmath $\sigma$},\mbox{\boldmath $\tau$})c(\mbox{\boldmath $\tau$},\mbox{\boldmath $\sigma$}) \big/ \sum_{\mbox{\boldmath $\sigma$},\mbox{\boldmath $\tau$}}c(\mbox{\boldmath $\sigma$},\mbox{\boldmath $\tau$})c(\mbox{\boldmath $\tau$},\mbox{\boldmath $\sigma$}) 
\end{align}
with $a=1,...,n$.

Inserting Eq. (21) into Eq. (17) and using Eqs. (3) and (5), we finally obtain in the large-$N$ limit
\be
\overline{{\cal N}(m,H)^n}\sim \exp\big[N \max_{\ss \mbox{c(\boldmath $\sigma$},\mbox{\boldmath $\tau$)},\mbox{\boldmath $g$}}\big\{\Lambda[c,\mbox{\boldmath $g$}]-m\sum_a g^a-\frac{z}{2}(\ln z+\ln \gamma)+\ln z!\big\}\big] \ .
\ee
\subsection{Annealed complexity}

We first consider the case $n=1$ corresponding to the annealed
average.  $\mbox{\boldmath $\sigma$}$ and $\mbox{\boldmath $\tau$}$ become standard Ising  variables $\sigma,\tau$ and the preceding equations simplify considerably. In particular, the function $c(\sigma,\tau)$ can only take four distinct values which we denote $c_{\scriptscriptstyle ++}$, $c_{\ss +-}$, $c_{\ss -+}$, and $c_{\ss --}$. Performing the integration over the random field distribution,  Eq. (18) becomes
\begin{align}
\Lambda[c,g]&=\ln\frac{\gamma^z}{z!}+\ln \sum_{\sigma,\tau_1...\tau_z=\pm 1}e^{g\sigma}c(\tau_1,\sigma)...c(\tau_z,\sigma) \frac{1}{2}\big[1+\mbox{erf}\big(\frac{[H+J\sum_1^z\tau_k]\sigma}{R\sqrt{2}}\big)\big]\nonumber\\
&=\ln \frac{\gamma^z}{z!}+\ln\big[e^g c_{\ss -+}^z  f_{\ss +}\big(\frac{c_{\ss ++}}{c_{\ss -+}}\big)+e^{-g} c_{\ss +-}^z f_{\ss -}\big(\frac{c_{\ss --}}{c_{\ss +-}}\big)\big]
\end{align}
where $\mbox{erf}(x)=(2/\sqrt{\pi})\int_0^x e^{-t^2} dt$ is the error function and
the functions $f_{\ss \pm}(x)$ are defined by
\be
f_{\ss\pm }(x)=\frac{1}{2}\sum_{n=0}^z\binom{z}{n}x^n\big[1+\mbox{erf}\Big(\frac{J(2n-z)\pm
H}{R\sqrt{2}}\Big)\big]\ .
\ee
Like in Ref.\cite{LD2001} we now change to the new variables $u=c_{\ss ++}/c_{\ss -+}$, $v=c_{\ss --}/c_{\ss +-}$, $t=c_{\ss +-}/c_{\ss -+}$, solve the saddle point equation for $c_{\ss -+}$, and substitute the result in Eq. (23). We then find that the dependence on $\gamma$ disappears (as it must be), and  the average number of metastable states with magnetization $m$ is given in the large-$N$ limit by $\overline{{\cal N}(m,H)}\sim \exp[\Sigma_A(m,H)\  N]$ with
\be
\Sigma_A(m,H)=\max_{u,v,t,g}\{-\frac{z}{2}\ln(u^2+t^2v^2+2t)+\ln[f_{\ss +}(u)e^g+t^zf_{\ss -}(v)e^{-g}] -gm\} \ .
\ee
Solving the stationarity condition with respect to $g$ gives
\be
g=\dfrac{1}{2}[z\ln t+\ln \frac{1+m}{1-m} -\ln \frac{f_{\ss +}(u)}{f_{\ss -}(v)} ] \ ,
\ee
and substituting into Eq. (26) we finally obtain
\begin{align}
\Sigma_A(m,H)=\max_{u,v,t}\{&-\frac{z}{2}\ln(u^2+t^2v^2+2t)+\frac{1+m}{2}\ln f_{\ss +}(u)  +\frac{1-m}{2}
\ln f_{\ss -}(v) \nonumber\\
&+z\frac{1-m}{2}\ln t -\frac{1+m}{2}\ln \frac{1+m}{2}-\frac{1-m}{2}\ln \frac{1-m}{2} \} \ .
\end{align}
The saddle point equations $\partial \Sigma_A/\partial v=0$ and $\partial \Sigma_A/\partial t=0$
may be solved analytically to give $t=t(v)=f'_-(v)/[zvf_{\ss -}(v)-v^2f'_-(v)]$
and $u=[(1+m)t^2v^2+2mt]^{1/2}/[1-m]^{1/2}$ (the negative branch of $u$ leads to  solutions that have no physical meaning). On the other hand, the
remaining equation, $\partial \Sigma_A/\partial u=0$, must be solved numerically.

The total complexity  $\Sigma_A^{max}(H)$ also  corresponds to the maximum of $\Sigma_A(m, H)$. It is given by Eq. (26) with $g=0$; one can associate to this maximum a value of the magnetization $m_A(H)$, which is then obtained by solving Eq. (27). One still has $t=t(v)$ but the saddle point equations for $u$ and $v$ must be solved numerically. For $H=0$ however, these equations admit the solution $u=v$ and one has to
solve the unique equation $t(v)=1$. As in the case of the pure ferromagnet\cite{LD2001}, this solution  corresponds to the fixed point of the transformation $u \to v$, $v \to u$, $t \to 1/t$ which leaves $\Sigma_A^{max}(H=0)$ invariant (when $R\neq 0$, this is the only solution of the saddle point equations). 

\subsection{Quenched complexity}

We now turn to the calculation of the quenched complexity  given by
\begin{align}
\Sigma_Q(m,H)=\lim_{N\to \infty}\frac{1}{N} \lim_{n\to 0}\frac{1}{n} \Big[\overline{{\cal N}(m,H)^n}-1\Big]
\end{align}
where, as usual, the order of the limits  $N\to \infty$ and  $n\to 0$ is inverted and the expression of $\overline{{\cal N}(m,H)^n}$ is extended to non-integer $n$ to allow the limit $n\to 0$ to be taken. Some ansatz concerning the form of the solution of the saddle point equation (19) as  $n\to 0$ is now required. As alluded to in the Introduction, a major difference between the ferromagnetic RFIM and spin-glass models comes from the role of frustration. For the Bethe lattice spin-glass\cite{MP2001}, frustration (due to the presence of large loops) induces a replica symmetry breaking instability similar to the one found in the fully-connected Sherrington-Kirkpatrick model and corresponding to the appearance of an exponential number of {\it pure} states. On the other hand, for the RFIM, this phenomenon does not occur in the infinite-range model\cite{SP1977}, nor on finite-connectivity lattices at the mean-field level (it may be checked for instance that the number of pure states on the Bethe lattice is indeed non-exponential\cite{F2003}). We thus assume that the solution of Eq. (19) remains invariant with respect to the permutation of the replica indices when $n\to 0$.  Setting $g^a=g$ in Eq. (23), we then obtain 
\begin{align}
\Sigma_Q(m,H)&=\max_{g}\{-mg+\lim_{n\rightarrow 0}\frac{1}{n}[\Lambda^*(g)-\frac{z}{2}(\ln z+\ln \gamma)+\ln z!]\}\nonumber\\
&=\Lambda_1^*(g^*)-mg^*
\end{align}
where $\Lambda^*(g)=\Lambda^*_0+n\Lambda^*_1(g)+O(n^2)$ is the value of $\Lambda$ at the saddle-point solution $c^*\mbox{(\boldmath $\sigma$},\mbox{\boldmath $\tau$)}$ and $g^*$ is the position of the maximum of $\Lambda^*(g)$. The second line of this equation also uses the relation $\Lambda^*_0=\frac{z}{2}(\ln z+\ln \gamma)-\ln z!$, which is a necessary condition for the $n\to 0$ limit to exist and be independent of $\gamma$. Note that Eq. (30) corresponds to a Legendre transformation between $\Sigma_Q$ and $\Lambda^*_1$, so that the stationnarity condition $\partial \Lambda_1^*/ \partial g^*=m$ implies that $\partial \Sigma_Q(m,H)/ \partial m=-g^*$.  In fact, as will be illustrated below, it is convenient to consider that both $\Sigma_Q$ and $m$ are parametrized by $g$ (to simplify the notations, the superscript * is dropped in the following), with $g$ going  from $-\infty$ to $+\infty$. One then has the overall symmetry $\Sigma_Q(-g,-H)=\Sigma_Q(g,H)$, $m(-g,-H)=m(g,H)$,   and the maximum number of metastable states is obtained by setting $g=0$.

The main consequence of replica-symmetry is that  $c(\mbox{\boldmath $\sigma$},\mbox{\boldmath $\tau$})$ is a function of the three variables $s=\sum_{a=1}^n\sigma^a$, $u=\sum_{a=1}^n\sigma^a\tau^a$, $t=\sum_{a=1}^n\tau^a$ only. This suggests, following Ref.\cite{BS2001}, to introduce an integral representation of $c(s,u,t)$ so as to  perform an analytic continuation of this function on the imaginary axis in the small $n$ limit where $s$, $u$, and $t$ may be treated as continuous variables. It turns out, however, that the situation is not that simple here because of the constraints on the values of $\sigma^a$ and $\tau^a$. To illustrate the problem, let us consider the case $z=2$. Eq. (19) then reads
\be
 c(\mbox{\boldmath $\sigma$},\mbox{\boldmath $\tau$})=\gamma e^{-\Lambda}\sum_{\mbox{\boldmath $\tau$}_1}c(\mbox{\boldmath $\tau$}_1,\mbox{\boldmath $\sigma$})F_2(\mbox{\boldmath $\sigma$},\mbox{\boldmath $\tau$},\mbox{\boldmath $\tau$}_1)
\ee
with
\be
F_2(\mbox{\boldmath $\sigma$},\mbox{\boldmath $\tau$},\mbox{\boldmath $\tau$}_1)=e^{g s}\int P(h)dh \prod_a \Theta[(H+h+J(\tau+\tau^a_1))\sigma^a] \ .
\ee
Performing the integration over the random field distribution in the four regions $h<-2J-H$, $-2J-H<h<-H$, $-H<h<2J-H$, and $2J-H<h$, we obtain
\begin{align}
F_2(\mbox{\boldmath $\sigma$},\mbox{\boldmath $\tau$},\mbox{\boldmath $\tau$}_1)&=
\frac{1-a}{2}e^{-gn}\prod_a\delta(\sigma^a;-1)+\frac{1-c}{2}e^{gn}\prod_a\delta(\sigma^a;1)\nonumber \\
&+\frac{a-b}{2}e^{g s}
\prod_a \big[\delta(\sigma^a;-1)
\delta(\tau^a_1;-\tau^a)+\delta(\tau_1^a;\tau^a)
\delta(\tau^a;\sigma^a)\big] \nonumber\\
&+\frac{b+c}{2}e^{g s}
\prod_a \big[\delta(\sigma^a;1)
\delta(\tau^a_1;-\tau^a)+\delta(\tau_1^a;\tau^a)
\delta(\tau^a;\sigma^a)\big]
\end{align}
where $a=\mbox{erf}((2J+H)/R\sqrt{2})$, $b=\mbox{erf}(H/R\sqrt{2})$, and $c=\mbox{erf}((2J-H)/R\sqrt{2})$. With this, Eq.(31) reads
\begin{align}
\frac{e^{\Lambda}}{\gamma}c(\mbox{\boldmath $\sigma$},\mbox{\boldmath $\tau$})&=\frac{1-a}{2}e^{-gn}\prod_a\delta(\sigma^a;-1)\sum_{\mbox{\boldmath $\tau$}_1}c(\mbox{\boldmath $\tau$}_1,-{\bf1})+\frac{1-c}{2}e^{gn}\prod_a\delta(\sigma^a;1)\sum_{\mbox{\boldmath $\tau$}_1}c(\mbox{\boldmath $\tau$}_1,{\bf1}) \nonumber \\
&+\frac{a-b}{2}e^{g s}\sum_{\mbox{\boldmath $\tau$}_1}c(\mbox{\boldmath $\tau$}_1,\mbox{\boldmath $\sigma$})
\prod_a \big[\delta(\sigma^a;-1)
\delta(\tau^a_1;-\tau^a)+\delta(\tau_1^a;\tau^a)
\delta(\tau^a;\sigma^a)\big] \nonumber\\
&+\frac{b+c}{2}e^{g s}\sum_{\mbox{\boldmath $\tau$}_1}c(\mbox{\boldmath $\tau$}_1,\mbox{\boldmath $\sigma$})
\prod_a \big[\delta(\sigma^a;1)
\delta(\tau^a_1;-\tau^a)+\delta(\tau_1^a;\tau^a)
\delta(\tau^a;\sigma^a)\big]
\end{align}
where $\pm{\bf1}=(\pm 1,...\pm 1)$. The difficulty comes from the presence of the Kronecker-$\delta$'s that precludes any straightforward use of a Fourier representation of $c(\mbox{\boldmath $\tau$},\mbox{\boldmath $\sigma$})$ in the limit $n\to 0$. Because of this, we work out the self-consistent equation (19) {\it before} taking the small $n$ limit, a nontrivial (if not unsolvable) task for a generic value of the connectivity. In fact, only the case $z=2$ seems to be simple enough to be handled analytically.  Inspection of  Eq. (34) suggests that the solution has the form
\begin{align}
\sqrt{\frac{\gamma}{2}}c(\mbox{\boldmath $\sigma$},\mbox{\boldmath $\tau$})&=A
\prod_a\delta(\sigma^a;-1)+Be^{g s}\prod_a\delta(\sigma^a;\tau^a)+ C
\prod_a \delta(\sigma^a;1)\nonumber \\
&+D(\mbox{\boldmath $\sigma$},\mbox{\boldmath $\tau$})\prod_a\big[\delta(\sigma^a;-1)\delta(\tau^a;1)+\delta(\tau^a;\sigma^a)\big] \nonumber\\
&+E(\mbox{\boldmath $\sigma$},\mbox{\boldmath $\tau$})\prod_a\big[\delta(\sigma^a;1)\delta(\tau^a;-1)+\delta(\tau^a;\sigma^a)\big] ,
\end{align}
which in turn, together with Eq. (21), yields a set of six coupled equations for the unknown quantities $\Lambda, A, B,C, D(\mbox{\boldmath $\sigma$},\mbox{\boldmath $\tau$}), E(\mbox{\boldmath $\sigma$},\mbox{\boldmath $\tau$})$. It is only at this stage that the $n\to 0$ limit may be taken, noticing that $D(\mbox{\boldmath $\sigma$},\mbox{\boldmath $\tau$})$ and $E(\mbox{\boldmath $\sigma$},\mbox{\boldmath $\tau$})$ are functions of two variables only (say, $s$ and $t$) instead of three because of the constraints on $\mbox{\boldmath $\sigma$}$ and $\mbox{\boldmath $\tau$}$ that are imposed by the Kronecker-$\delta$ (it is easy to check that the third variable $u=n-\vert s-t\vert$). We then introduce the integral representations
\be
D(s,t)=\int dx dy {\hat D}(x,y) e^{xs +yt} \ , \ E(s,t)=\int dx dy {\hat E}(x,y) e^{xs +yt}
\ee
where  ${\hat D}(x,y)$ and ${\hat E}(x,y)$ are the Fourier transforms of $D(s,t)$ and $E(s,t)$ when $n\to 0$,
\be
{\hat D}(x,y)=\int \frac{ds}{2\pi} \frac{dt}{2\pi}  D(is,it)e^{-i(sx+ty)}, \ {\hat E}(x,y)=\int \frac{ds}{2\pi} \frac{dt}{2\pi}  E(is,it)e^{-i(sx+ty)} \ ,
\ee
and we expand all quantities at order $n$ (i.e. $A=A_0+nA_1+O(n^2), ...
,{\hat E}(x,y)={\hat E}_0(x,y)+n{\hat E}_1(x,y)+O(n^2)$) as detailed in Appendix B. This allows one to solve the equations at the first two orders in $n$ and, after some lengthy algebra, to compute $\Lambda_1(g)$ (see Eq. B25) from which we finally get

\be
\Sigma_Q(g,H)=\frac{1}{2}[1+b+\frac{(1-a)(b+c)}{a-b}][S_3(g)-g\frac{\partial S_3(g)}{\partial g}]
\ee
where $S_3(g)=\int dx dy {\hat D}_0(x,y)\ln[2\cosh x]$. A closed expression of $m(g,H)$ (Eq. B27) may be also obtained from Eq. (22), but it contains more complicated integrals of ${\hat D}_0(x,y)$ and it is in fact more convenient to compute the magnetization  by numerically differentiating the expression (B25) of $\Lambda_1(g)$ with respect to $g$.

The only remaining task consists in solving numerically the coupled integral equations satisfied by ${\hat D}_0(x,y)$ and ${\hat E}_0(x,y)$. This is outlined in Appendix C.

\section{Results for the one-dimensional chain ($z=2$)}

When $z=2$ the Bethe lattice is just a collection of disconnected closed loops. In
the thermodynamic limit, the dominant contributions to self-averaging quantities come from the loops of macroscopic size, and one thus expects the results to be the same as for the one-dimensional RFIM. Although no phase transitions  occur in one dimension, frustration makes the problem nontrivial, and the relation between the shape of the hysteresis loop and the distribution of the metastable states can be fully elucidated.
Note that the typical number of metastable states of the random-field Ising chain and their  distribution in magnetization and energy were computed a few years ago by a numerical transfer-matrix method\cite{MJWS1993}, but the study was restricted to the case of  zero external field: therefore, the relation with the $T=0$ hysteretic behavior was not investigated. Moreover, some results of Ref.\cite{MJWS1993} appear to violate the convexity inequality $\overline{\ln {\cal N}}\leq \ln\overline{{\cal N}}$, as will be shown in the following.

\begin{figure}
\begin{center}
\resizebox{9cm}{!}{\includegraphics*{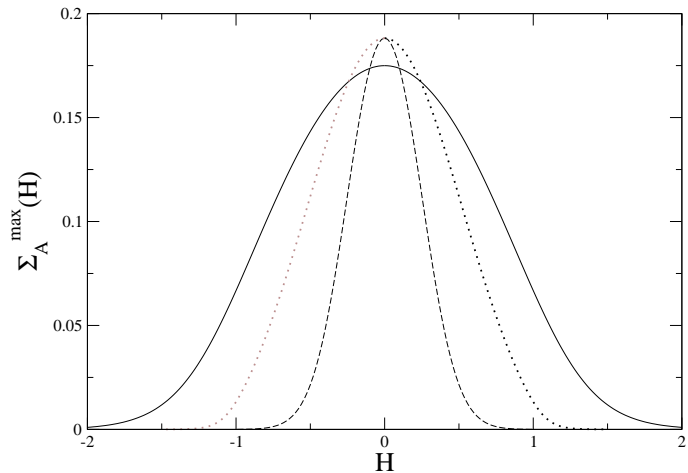}}
\caption{Total annealed complexity  $\Sigma_A^{max}(H)$ for $z=2$ and $R=1$ (solid line), $0.5$ (dotted line) and $0.25$ (dashed line). }
\end{center}
\end{figure}

\subsection{Annealed complexity}

Let us first consider the total annealed complexity  $\Sigma_A^{max}(H)$ as obtained from Eq. (26) with $g=0$. Its dependence on $H$ is illustrated in Fig. 1 for different values of the disorder strength $R$ (hereafter we take $J=1$ in the numerical applications). It can be seen that the presence of the magnetic field reduces the total number of metastable states, in agreement with the expectation that a sufficiently large field imposes the existence of a unique state. As $R$ decreases, the metastable states concentrate in the vicinity of $H=0$ and in the limit $R \to 0$ they are completely destroyed by an infinitesimal field. From the solution of the equation $t(v)=1$,  the zero-field complexity is given by 
\be 
\Sigma_A^{max}(H=0)=\ln\big[\frac{1}{2}+\frac{1}{2}\sqrt{1+\mbox{erf}^2(\frac{J\sqrt{2}}{R}})\big] \ ,
\ee
in agreement with a recent transfer matrix calculation\cite{BM2004}. It may be noted that this quantity  approaches $\ln \frac{1+\sqrt{2}}{2}\approx 0.188$ when $R\to 0$, a value much smaller than the complexity  of the pure ferromagnet, $\Sigma^{max}_{R=0}(H=0)=\ln \frac{1+\sqrt{5}}{2}\approx 0.481$\cite{LD2001,D2000}. This singular behavior of the zero-field complexity computed from Eqs. (25-28) in the limit $R \to 0$ comes from the fact that marginal states, absent when $R\neq 0$, are not properly counted when $R=0$: indeed, $f(x)=[1+\mbox{erf}(x/R\sqrt{2})]/2\to \Theta(x)$ for $x\neq0$ whereas $f(0)=1/2$; to also include in our calculation the marginal states that exist in the pure system, one should  instead set $\Theta(0)=1$ as in Ref.\cite{LD2001}. Actually, all metastable states are marginal in the Ising chain at zero field  (they simply correspond to different positions of the walls separating domains of up and down spins) and one can directly calculate a ``biased"  complexity  by associating to each  domain wall an arbitrary weight $p$ with $0\leq p\leq1$; the result is  $\ln \frac{1+\sqrt {1+4p^2}}{2}$,  which  for $ p=1/2$ identifies with the complexity found in the $R\rightarrow 0$ limit of the present approach.

\begin{figure}
\begin{center}
\resizebox{9cm}{!}{\includegraphics*{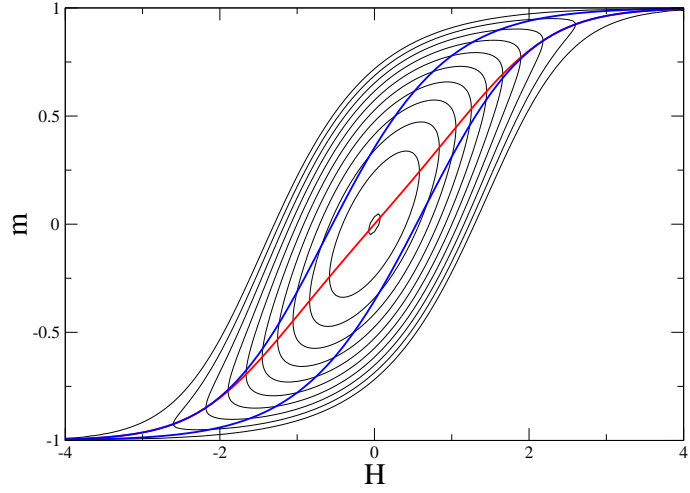}}
\caption{(Color online) Projection of $\Sigma_A(m,H)$ onto the $H-m$ plane for $z=2$ and $R=2$. The contours are separated by $10^{-2}$ (with $\Sigma_A(0,0)\approx 0.100$). The red curve is the locus $m_A(H)$ of the maxima of $\Sigma_A(m,H)$. The blue curve is the $T=0$ saturation hysteresis loop calculated in Ref.\cite{S1996}. }
\end{center}
\end{figure}

In Fig. 2, we plot the projection of the annealed complexity with fixed magnetization
$\Sigma_A(m,H)$ onto the $H-m$ plane for $R=2$. Only the region where the complexity is positive
is shown: this is the region where the number of metastable states is
exponentially large and thus comparable to the total number of states
in the system; in the rest of the plane, $\Sigma_A(m,H)<0$, and the number of metastable states is equal to zero in the thermodynamic limit.
Several curves in Fig. 2 are of special interest: i) the  contour
$\Sigma_A(m,H)=0$, ii) the curve $m_A(H)$ which is the locus of the maxima of $\Sigma_A(m,H)$ and thus gives the magnetization of
the metastable states dominating the (unrestricted) average number $\overline{{\cal N}(H)}$
in the thermodynamic limit (i.e. $\Sigma_A^{max}(H)\equiv\Sigma_A(m_A(H),H)$), and iii) the $T=0$ saturation hysteresis loop calculated in Ref.\cite{S1996} from probabilistic arguments.
We see on this example that the contour $\Sigma_A(m,H)=0$ significantly overestimates the size of the actual hysteresis loop (a similar observation was made in
the case of the Sherrington-Kirkpatrick model\cite{PZZ1999}). 
The fact that the saturation loop lies well inside the contour $\Sigma_A(m,H)=0$ 
can be understood as follows. Each realization  of the disorder has a different saturation loop and, as shown in Appendix A, there are no metastable states outside that loop. However, the observed saturation loop is obtained by averaging the magnetization over all disorder realizations, an averaging which is dominated by the {\it typical} realizations and as such is related to the quenched complexity $\Sigma_Q(m,H)$ that from Jensen's inequality is always less than $\Sigma_A(m,H)$. So, metastable states do exist outside the (average) saturation loop, there may even exist an exponentially large number of them, but these states are {\it non-typical} and essentially non-observable.

\subsection{Quenched complexity}

\begin{figure}
\begin{center}
\resizebox{9cm}{!}{\includegraphics*{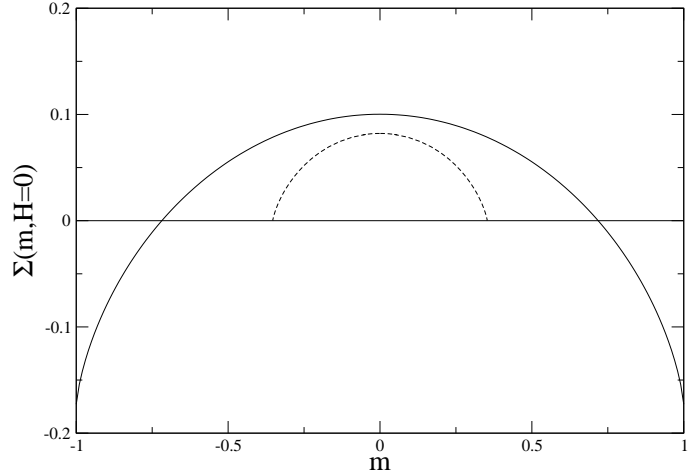}}
\caption{Annealed (solid line) and quenched (dashed line) complexities in zero external field for $z=2$ and $R=2$ as a function of magnetization.}
\end{center}
\end{figure}
As discussed in section 3 and Appendix B, the quenched complexity and the magnetization are parametrized by $g$ (see, e.g., Eq. (38)), with the maximum number of metastable states corresponding to $g=0$. As $g$ goes from $0$ to $+\infty$ (resp. $-\infty$), $\Sigma_Q(g,H)$ decreases monotonically towards $0$ whereas $m(g,H)$ increases (resp. decreases) monotonically,  the surface $\Sigma_Q(m,H)$ having roughly the same overall shape as $\Sigma_A(m,H)$: this is illustrated in Fig. 3 for $R=2$. Note however that the derivative $\partial \Sigma(m,H)/ \partial m$, i.e., the value of $g$,  becomes infinite for nontrivial values of the magnetization for the quenched complexity whereas this only occurs for $m=+1$ or $-1$ for the annealed complexity.

The projection of $\Sigma_Q(m,H)$
onto the $H-m$ plane is shown in Fig. 4 for $R=2$, together with $m(g=0,H)$, the typical magnetization of the metastable states\cite{note4}. The region where $\Sigma_Q(m,H)$ is positive is clearly thinner than the corresponding region for $\Sigma_A(m,H)$ shown in Fig. 2.
In fact, as proven analytically in Appendix C, the contour $\Sigma_Q(m,H)=0$ obtained in the limit $g\to \pm \infty$ corresponds  {\it exactly} to the saturation loop calculated in Ref.\cite{S1996}.  This means that the $T=0$  hysteresis loop is indeed the enveloppe of {\it all} the typical metastable states in the thermodynamic limit (i.e. the entropy density is strictly positive everywhere inside the loop).

\begin{figure}
\begin{center}
\resizebox{9cm}{!}{\includegraphics*{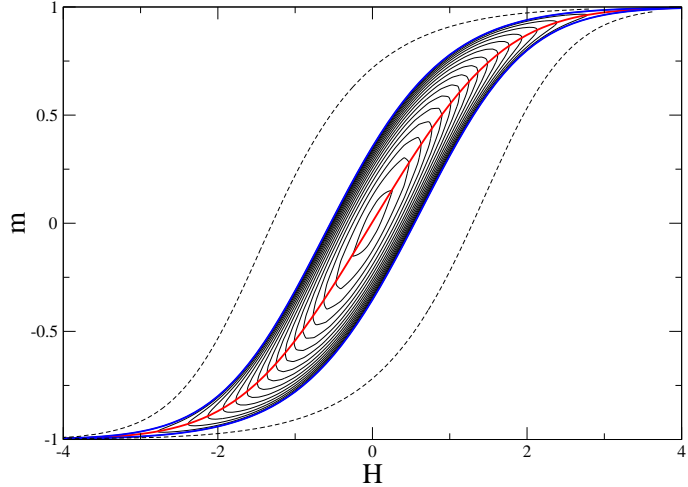}}
\caption{(Color online) Projection of $\Sigma_Q(m,H)$ onto the $H-m$ plane for $z=2$ and $R=2$. The contours are separated by $5.10^{-3}$ (with $\Sigma_Q(0,0)\approx 0.082$). The contour $\Sigma_Q(m,H)=0$ (in blue) coincides with the saturation hysteresis loop calculated in Ref.\cite{S1996}, as shown in Appendix C. The dashed curve is the annealed contour $\Sigma_A(m,H)=0$. The red curve is the typical magnetization of the metastable states (the locus of the maxima of the quenched complexity). }
\end{center}
\end{figure}
We show in Fig. 5 the total quenched complexity in zero field $\Sigma_Q^{max}(H=0)$, which is equal to $\Sigma_Q(m=0,H=0)$, as a function of $R$.  It can be seen that the curve is very flat near $R=0$ and that $\lim_{R\to 0}\Sigma_Q^{max}(H=0)\approx0.179$ (see Appendix C), which differs from the limit of the annealed complexity $\lim_{R\to 0}\Sigma_A^{max}(H=0)=\ln \frac{1+\sqrt{2}}{2}\approx 0.188$ (the difference is well outside the error bars in the numerics), itself smaller than the total complexity of the pure system. As discussed above in commenting the latter result, those differences  come from the singular features of the limit $R\to 0$: when $R\neq 0$, no metastable states are marginal, whereas when $R=0$ all metastable states are marginal. However, contrary to what happens for the annealed complexity in the limit $R\to 0$, the present procedure for calculating the quenched complexity does not reduce  to a simple (and biased) way of counting the marginal states in the pure system. 
(Note also that $\Sigma_Q^{max}(H=0)\approx0.176$ for $R=0.75$, which is at odds with the value $\Sigma_Q^{max}(H=0)\approx 0.25$ that can be extracted from the Figure 5 of Ref.\cite{MJWS1993}; this latter result is actually larger than the corresponding annealed complexity obtained from Eq. (39), $\Sigma_A^{max}(H=0)\approx 0.186$, which clearly violates Jensen's inequality and suggests some flaw in the calculations of Ref.\cite{MJWS1993}.)
\begin{figure}
\begin{center}
\resizebox{8cm}{!}{\includegraphics*{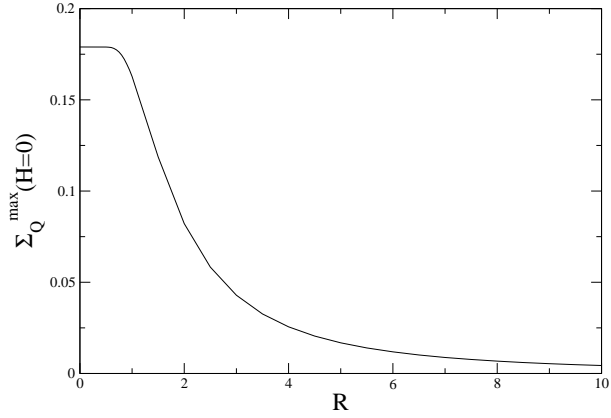}}
\caption{Total quenched complexity  in zero field $\Sigma_Q^{max}(H=0)$ for $z=2$ as a function of disorder strength.}
\end{center}
\end{figure}

\section{Annealed complexity for $z>2$}

\begin{figure}
\begin{center}
\resizebox{11cm}{!}{\includegraphics*{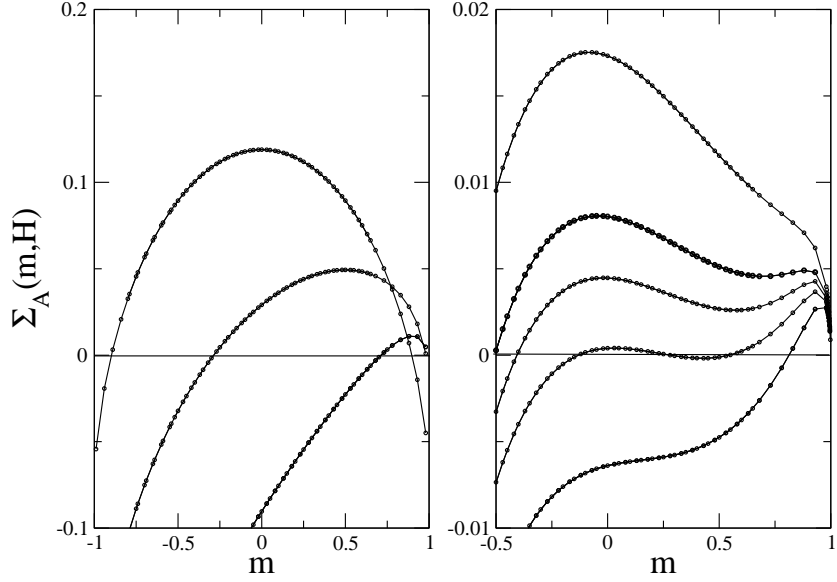}}
\end{center}
\caption{Annealed complexity  $\Sigma_A(m,H)$ for $z=3$ as a function of $m$ for different values of $H$. Left panel: $R=2$; from top to bottom $H=0, 1.30,2$. Right panel: $R=1.2$; from top to bottom $H=1.30,1.34,1.355,1.372,1.40$ (the  values of $\Sigma_A(m=1,H)$ not visible here are slightly negative).}
\end{figure}
\begin{figure}
\begin{center}
\resizebox{9cm}{!}{\includegraphics*{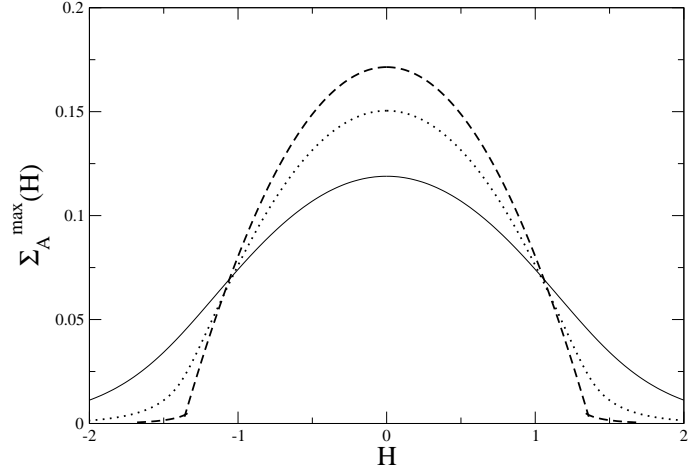}}
\caption{Total annealed complexity  $\Sigma_A^{max}(H)$ for $z=3$ and $R=2$ (solid line), $1.5$ (dotted line), and $1.2$ (dashed line). Note the discontinuity in the slope of $\Sigma_A^{max}(H)$ at $H\approx 1.355$ for $R=1.2$.}
\end{center}
\end{figure}
For $z>2$ one expects new phenomena to take place. In particular, the complexity may not be anymore a convex function of $H$ and $m$ and phase transitions may occur.

Let us first consider the case $z=3$. In Fig. 6 we show $\Sigma_A(m,H)$ as a function of the magnetization for two different values of $R$ and several values of the external field (we only consider positive values of $H$ because of the symmetry $\Sigma(m,-H)=\Sigma(-m,H)$). As can be seen in the left panel of the figure, the complexity is always a convex function of the magnetization when $R$ is large enough. As a consequence, the total complexity $\Sigma_A^{max}(H)=\max_m\Sigma_A(m,H)$ is a continuous function of $H$, as shown in Fig. 7. This is also the case for the corresponding magnetization $m_A(H)$. There are only two values of $m$ for which $\Sigma_A(m,H)=0$ (note that the two limiting values $\lim_{m \rightarrow \pm 1}\Sigma_A(m,H)=\ln  [\big (1+\mbox{erf}[(zJ\pm H)/(R\sqrt{2})]\big)/2]$ are negative), and the overall surface $\Sigma_A(m,H)$ has roughly the same shape as for $z=2$ (the projection onto the $H-m$ plane resembles the one shown in Fig. 2). 

The behavior is more complicated at low disorder as illustrated by the curves in the right panel of Fig. 6. For $R$ smaller than some critical value $(R\lesssim1.28)$ and in a certain range of $H$, $\Sigma_A(m,H)$ may indeed display two positive maxima at two distinct values of $m$. As a consequence, there is a discontinuity in $\partial \Sigma_A^{max}(H)/ \partial H$ at some value of the field ($H\approx 1.355$ for $R=1.2$) as shown in Fig. 7 and, accordingly, there is a jump in $m_A(H)$. Moreover, for $R$ even smaller, there is a small range of $H$ for which  $\Sigma_A(m,H)=0$ for four distinct values of the magnetization. This yields an overall surface $\Sigma_A(m,H)$ that has a more complicated shape than in the strong-disorder regime, as illustrated by the projection onto the $H-m$ plane shown in Fig. 8.  In particular, one can observe in the right pannel the ``S" shape of the  contour $\Sigma_A(m,H)=0$ in the range $1.3<H<1.4$. Note that the actual hysteresis loop calculated in Ref.\cite{DSS1997} is again inside the contour $\Sigma_A(m,H)=0$ as it must be. The fact that $\partial m_A(H)/ \partial H$ is negative in a certain range of $H$ is  surprising and can only be explained by the existence of metastable states that have a quite  non-typical magnetization. One must recall, however, that $m_A(H)$ is not an observable quantity.   For the same reason, the non-monotonic behavior associated with the contour $\Sigma_A(m,H)=0$ bears no connection with the behavior of the actual hysteresis loop: this latter does not display any finite jump nor phase transition for $z=3$\cite{DSS1997}.

\begin{figure}[hbt]
\begin{center}
\resizebox{12cm}{!}{\includegraphics*{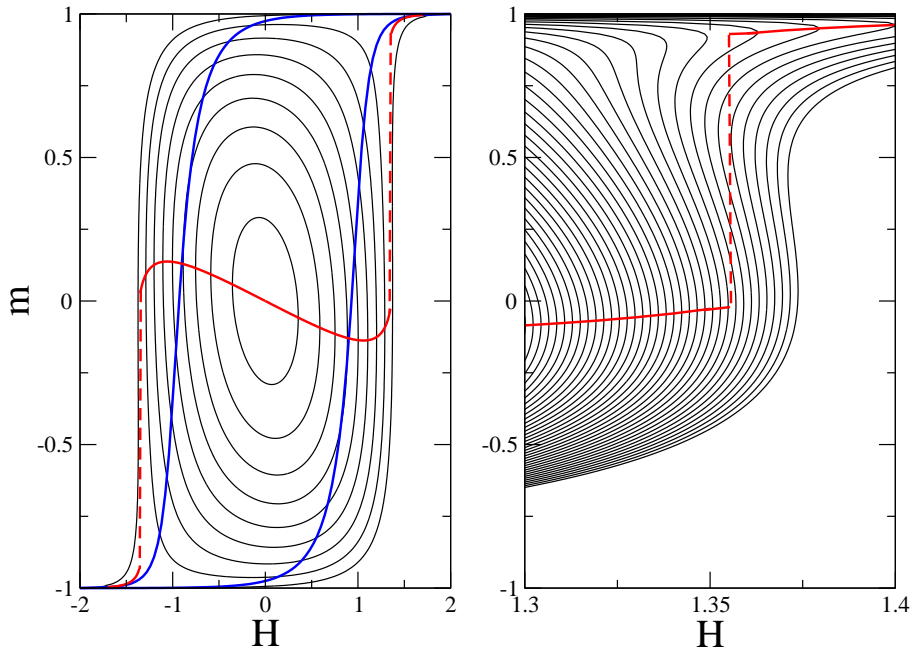}}
\end{center}
\caption{(Color on line) Projection of  $\Sigma_A(m,H)$ onto the $H-m$ plane for $z=3$ and $R=1.2$. In the left panel, the contours are separated by $2.10^{-2}$ (with $\Sigma_A(0,0)\approx 0.171$). The red curve is the locus $m_A(H)$ of the maxima of $\Sigma_A(m,H)$. The blue curve is the $T=0$ hysteresis loop calculated in Ref.\cite{DSS1997}. The right pannel is a close-up of the figure in the range $1.3 <H<1.4$, with a spacing $5.10^{-4}$ between the contours.}
\end{figure}
\begin{figure}[hbt]
\begin{center}
\resizebox{7cm}{!}{\includegraphics*{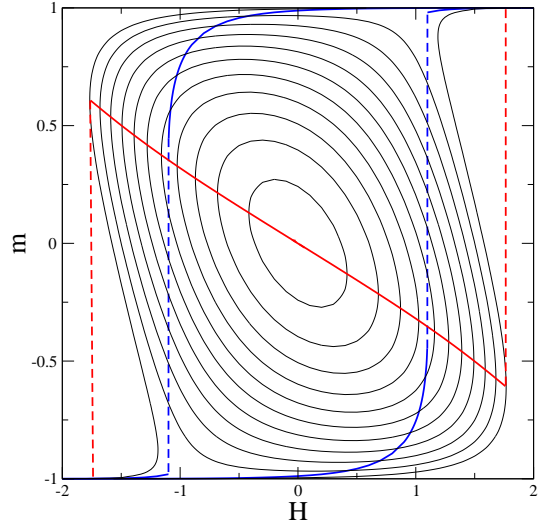}}
\end{center}
\caption{(Color on line) Projection of  $\Sigma_A(m,H)$ onto the $H-m$ plane for $z=4$ and $R=1.5$. The contours are separated by $1.5.10^{-2}$ (with $\Sigma_A(0,0)\approx 0.159$). The red curve is the locus $m_A(H)$ of the maxima of $\Sigma_A(m,H)$. The blue curve is the $T=0$ hysteresis loop calculated in Ref.\cite{DSS1997}.}
\end{figure}

For $z=4$, the situation changes for the hysteresis: a phase transition and a jump discontinuity appear for $R\lesssim 1.78$\cite{DSS1997}. $\Sigma_A(m,H)$, on the other hand, has the same type of behavior as for $z=3$. In particular, for $R\lesssim 2.25$ and in a certain range of the external field $H$, it has two positive maxima as a function of $m$ and four zeros, like in the left panel of Fig. 6. The projection of $\Sigma_A(m,H)$ onto the $H-m$ plane is shown in Fig. 9 for $R=1.5$. $m_A(H)$ has again a strange behavior,
first decreasing (resp. increasing) as $H$ is increased (resp. decreased) and then jumping to a value which is very close to $1$ (resp. $-1$). This is probably the generic behavior for $z\geq 4$.

\section{Discussion}

\begin{figure}[hbt]
\begin{center}
\resizebox{8cm}{!}{\includegraphics*{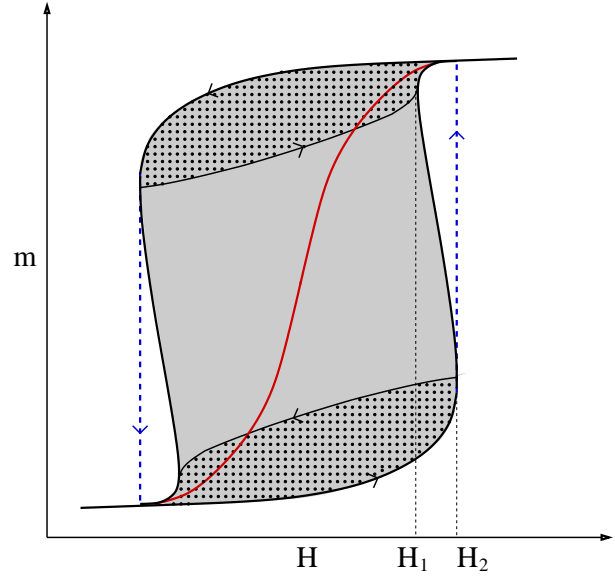}}
\end{center}
\caption{(Color on line) Speculative distribution of the metastable and H-states in the $H-m$ plane for $z\geq 4$ in the weak-disorder regime where the hysteresis loop has a jump at $H=\pm H_2$. The number of metastable states is exponentially large in system size in the shaded region and zero outside. Dots indicate the region where the number of H-states is exponentially large. The red curve is the typical magnetization of the metastable states.}
\end{figure}

One important conclusion of the computations presented above is that the annealed approximation does not properly describe the actual distribution of the metastable states of the RFIM in the field-magnetization diagram (i.e., the distribution of metastable states associated with {\it typical} realizations of the disorder). The calculations are done here on a Bethe lattice, but this conclusion is probably true for any lattice. Predictions can even be qualitatively  wrong, as illustrated by the fact that a phase transition  corresponding to a change of convexity of the annealed complexity occurs for $z=3$ whereas the typical behavior changes for $z\geq 4$ only. On the other hand, the quenched calculation for $z=2$ shows that the entropy  density associated to the  metastable states is strictly positive everywhere inside the saturation hysteresis loop. It is very likely to be always the case in the strong-disorder regime where the saturation loop is continuous. The situation is less clear for $z\geq 4$  in the weak-disorder regime where the saturation loop displays a jump discontinuity in the magnetization. Since we have not  so far been able to carry out the quenched calculation to the end, we can only speculate about the actual behavior of the quenched complexity $\Sigma_Q(m,H)$. A possible scenario is depicted in Fig. 10.
The rationale for this scenario is very intuitive: we simply assume that the jump in the magnetization curve at some field $H_2(R)$ (or, symmetrically, at $- H_2(R)$) occurs because, with probability $1$ in the thermodynamic limit, there are no metastable states with a low magnetization for $H>H_2$. The ``S" shape of the curve $\Sigma_Q(m,H)=0$ is speculative but it is tempting to associate this contour with  the real but ``unphysical" roots of the self-consistent polynomial equation for the fixed-point probability $P^*$ derived in Ref.\cite{DSS1997} to describe the saturation hysteresis loop.  Let us recall that for $z=4$ this equation has three real solutions in the range  $H_1<H<H_2$ with two of them merging and becoming complex at $H=H_1$ and $H=H_2$. Of course, one cannot discard a more complicated scenario to occur, in particular when there are more than three roots for $z>4$. In principle, 
there is also no objection for the existence of a jump discontinuity  in the typical magnetization, i.e., the magnetization of the states corresponding to $\max_m\Sigma_Q(m,H)$ (which is not the case considered in Fig. 10). One could indeed imagine a mechanism similar to the one displayed in Fig. 6 with $\Sigma_Q(m,H)$ having two maxima as a function of $m$ in a certain range of $H$: a ``first-order" transition could then occur when the two maxima have the same height.  These are interesting open questions which 
remain to be resolved before the problem of the distribution of metastable states of the RFIM on the Bethe lattice can be considered as fully understood. It would be of course even more interesting  to know the actual behavior on a $3$-d lattice, but this can only be done numerically.

Finally, let us remark that very little is known about a  special category of metastable states, the so-called ``$H$-states"\cite{BM2004b}, which are the spin configurations that can be reached by a field history starting from one of the saturated states (these configurations are characterized by the  applied sequence of reversal fields $\{H_i\}$). The $H$-states are thus experimentally observable by recording various hysteresis loops and subloops.  The most easily accessible ones, which we have discussed in this paper, are those associated with the saturation hysteresis loop.
The $H$-states are believed to be low-energy configurations, and it has been recently suggested that their number also increases exponentially with the system size\cite{BM2004b}. An analytical check of this result would be much welcome, even in one dimension.  As was pointed out in introduction, in the weak-disorder regime, for $z\geq 4$, a region around the origin in the $(H-m)$ plane becomes inaccessible to any field history, which forbids a demagnetization process to be completed\cite{CGZ2002}.  The corresponding distribution of the $H$-states is as illustrated in Fig. 10. In particular, the probability of finding $H$-states in the central part of the  loop (the region of the $(H-m)$ plane without dots) is zero in the thermodynamic limit. This is related to the fact that the states on the concave parts of the envelope of the shaded region are not reachable by any field history because of the jump in the major hysteresis loop\cite{note10}.

\acknowledgments
We acknowledge useful discussions with V. Basso, S. Zapperi, and E. Vives.  
The Laboratoire de Physique Th\'eorique des Liquides is the UMR 7600 of
the CNRS.

\appendix
\section{}
In this Appendix, we show that the probability to find metastable states outside the saturation hysteresis loop vanishes in the thermodynamic limit when the interactions are ferromagnetic. This results from the fact that for any given realization of the random fields, there can be no metastable states outside the corresponding saturation hysteresis loop. Indeed, the properties of the $T=0$ single-spin-flip dynamics are such that for any given value of the magnetic field $H$, all sequences starting from the fully polarized configuration (with all spins up or all spins down), sequences in which unstable spins are flipped, lead to the same metastable state on the upper (resp. lower) branch of the hysteresis loop\cite{S1993}. This is true in particular for a parallel updating of all unstable spins. 
Let us denote $\cal{C}_{\alpha}(H)$, $\alpha=1,...,p$, the $p$ configurations obtained after each updating steps along this process with ${\cal C}_0(H)$ being the initial fully polarized state with, say, all spins up and ${\cal C}_{p}(H)= {\cal C}_+(H)$ the final state on the upper branch of the hysteresis loop. The magnetization of course decreases as $\alpha$ increases. 
 Consider now a metastable configuration $\cal{C}(H)$. Because of the ferromagnetic nature of the couplings, the unstable spins that flip from ${\cal C}_{0}$ to ${\cal C}_{1}$ must be down in ${\cal C}(H)$; the same is true for the unstable spins that flip between ${\cal C}_{1}$ and ${\cal C}_{2}$ and the reasoning  can be repeated so that all down spins in ${\cal C}_p(H)={\cal C}_+(H)$ must also be down in $\cal{C}(H)$. The magnetization of $\cal{C}(H)$ is thus less than that of 
${\cal C}_+(H)$. This is a form of the ``no-passing rule"\cite{S1993}. The argument also applies when starting from the fully polarized state with all spins down, which completes the demonstration. As the property is true for all realizations of the disorder, this applies in particular to the ``typical" realizations whose number dominates the whole distribution, so that one expects no ``typical" metastable states outside the (average) saturation hysteresis loop.

\section{}

In this Appendix, we detail the calculation of the quenched complexity for $z=2$.

Injecting (35) into Eq. (34), we get the following equations,
\begin{subequations}
\begin{align}
\frac{e^{\Lambda}}{\gamma}A=\frac{1-a}{2}e^{-gn}\big[A+Be^{-gn}+C+D(-{\bf 1},-{\bf 1})+\sum_{\mbox{\boldmath $\tau$}_1}E(\mbox{\boldmath $\tau$}_1,-\mbox{\boldmath $1$})\big]+\frac{a-b}{2}e^{-gn}A
\end{align}
\begin{align}
\frac{e^{\Lambda}}{\gamma}C=\frac{1-c}{2}e^{-gn}\big[A+Be^{-gn}+C+E({\bf 1},{\bf 1})+\sum_{\mbox{\boldmath $\tau$}_1}D(\mbox{\boldmath $\tau$}_1,\mbox{\boldmath $1$})\big]+\frac{b+c}{2}e^{-gn}C
\end{align}
\be
\frac{e^{\Lambda}}{\gamma}B=\frac{a-b}{2}C+\frac{b+c}{2}A
\ee
\end{subequations}
and
\begin{subequations}
\begin{align}
\frac{e^{\Lambda}}{\gamma}D(\mbox{\boldmath $\sigma$},\mbox{\boldmath $\tau$})&=\frac{a-b}{2}e^{g s}\Big\{Be^{g s}+D(\mbox{\boldmath $\sigma$},\mbox{\boldmath $\sigma$})+\sum_{\mbox{\boldmath $\tau$}_1}E(\mbox{\boldmath $\tau$}_1,\mbox{\boldmath $\sigma$})\nonumber\\
&\times\prod_a \big[\delta(\sigma^a;-1)
\delta(\tau^a_1;-\tau^a)+\delta(\tau_1^a;\tau^a)
\delta(\tau^a;\sigma^a)\big]\Big\}
\end{align}
\begin{align}
\frac{e^{\Lambda}}{\gamma}E(\mbox{\boldmath $\sigma$},\mbox{\boldmath $\tau$})&=\frac{b+c}{2}e^{g s}\Big\{Be^{g s}+E(\mbox{\boldmath $\sigma$},\mbox{\boldmath $\sigma$})+\sum_{\mbox{\boldmath $\tau$}_1}D(\mbox{\boldmath $\tau$}_1,\mbox{\boldmath $\sigma$})\nonumber\\
&\times\prod_a \big[\delta(\sigma^a;1)
\delta(\tau^a_1;-\tau^a)+\delta(\tau_1^a;\tau^a)
\delta(\tau^a;\sigma^a)\big]\Big\} \ .
\end{align}
\end{subequations}
After some algebra we also obtain from Eq. (21) the additional equation
\begin{align}
&(A+C)^2+2B(Ae^{-gn}+Ce^{gn})+B^2(2\cosh 2g)^n +2B\sum_{\mbox{\boldmath $\sigma$}}e^{gs}[D(\mbox{\boldmath $\sigma$},\mbox{\boldmath $\sigma$})+E(\mbox{\boldmath $\sigma$},\mbox{\boldmath $\sigma$})]\nonumber\\
&+2A[D(-{\bf 1},-{\bf 1})+\sum_{\mbox{\boldmath $\tau$}}E(\mbox{\boldmath $\tau$},-{\bf 1})]+2C[E({\bf 1},{\bf 1})+\sum_{\mbox{\boldmath $\tau$}}D(\mbox{\boldmath $\tau$},{\bf 1})]\nonumber\\
&+\sum_{\mbox{\boldmath $\sigma$}}[D^2(\mbox{\boldmath $\sigma$},\mbox{\boldmath $\sigma$})+E^2(\mbox{\boldmath $\sigma$},\mbox{\boldmath $\sigma$})]+2\sum_{\mbox{\boldmath $\sigma$},\mbox{\boldmath $\tau$}}D(\mbox{\boldmath $\sigma$},\mbox{\boldmath $\tau$})E(\mbox{\boldmath $\tau$},\mbox{\boldmath $\sigma$})\prod_a[\delta(\sigma^a;1)\delta(\tau^a;-1)+\delta(\sigma^a;\tau^a)]=1
\end{align}
Eqs. (B1-B3) constitute a set of six coupled equations that must be solved at the first two orders in $n$.
As already mentioned after Eq. (35), it is important to note that $D(\mbox{\boldmath $\sigma$},\mbox{\boldmath $\tau$})$ and $E(\mbox{\boldmath $\sigma$},\mbox{\boldmath $\tau$})$  are functions of two variables only. Indeed, the term  $\prod_a\big[\delta(\sigma^a;-1))\delta(\tau^a;1)+\delta(\tau^a;\sigma^a)\big]$ in Eq. (36) imposes that $\sigma^a=1$ and $\tau^a=-1$ cannot hold simultaneously. This in turns implies that $u=n-\vert s-t\vert$.

We now expand both sides of Eqs. (B1-B3) at order $n$.
Let us first consider objects like $D({\bf -1},{\bf -1}),\sum_{\mbox{\boldmath $\tau$}_1}D(\mbox{\boldmath $\tau$}_1,\mbox{\boldmath $1$})$, etc... in Eqs. (B1).
Using the integral representations  (37) of  $D(\mbox{\boldmath $\sigma$},\mbox{\boldmath $\tau$})$ and $E(\mbox{\boldmath $\sigma$},\mbox{\boldmath $\tau$})$, we find
\begin{align}
D({\bf -1},{\bf -1})&=\int dx dy {\hat D}(x,y) e^{-n(x+y)}\nonumber\\
&=D_0+n(D_1-S_1-S_2)+O(n^2)\end{align}
\begin{align}
\sum_{\mbox{\boldmath $\tau$}_1}D(\mbox{\boldmath $\tau$}_1,\mbox{\boldmath $1$})&=
\int dx dy {\hat D}(x,y) e^{ny}\sum_{\mbox{\boldmath $\tau$}_1} e^{x\sum_a\tau_1^a}\nonumber\\
&=D_0+n(D_1+S_2+S_3)+O(n^2)
\end{align}
where $D_0= \int dx dy {\hat D}_0(x,y)$, $D_1=\int dx dy {\hat D}_1(x,y)$, $S_1=\int dx dy x{\hat D}_0(x,y)$, $S_2=\int dx dy y{\hat D}_0(x,y)$, and $S_3=\int dx dy {\hat D}_0(x,y)\ln[2\cosh x]$. (Similar expressions are obtained for $E({\bf 1},{\bf 1})$ and $\sum_{\mbox{\boldmath $\tau$}_1}E(\mbox{\boldmath $\tau$}_1,\mbox{-\boldmath $1$})$, with $E_0,E_1,S'_1,S'_2$ and $S'_3$ replacing respectively $D_0,D_1,-S_1,-S_2$ and $S_3$.)

In the same way, one has in Eq. (B3)
\begin{align}
\sum_{\mbox{\boldmath $\sigma$}}e^{gs}[D(\mbox{\boldmath $\sigma$},\mbox{\boldmath $\sigma$})+E(\mbox{\boldmath $\sigma$},\mbox{\boldmath $\sigma$})]&=
\int dx dy [{\hat D}(x,y)+{\hat E}(x,y)] \sum_{\mbox{\boldmath $\sigma$}} e^{s(g+x+y)}\nonumber\\
&=D_0+E_0+n(D_1+E_1+S_4)+O(n^2)
\end{align}
\begin{align}
\sum_{\mbox{\boldmath $\sigma$}}[D^2(\mbox{\boldmath $\sigma$},\mbox{\boldmath $\sigma$})+E^2(\mbox{\boldmath $\sigma$},\mbox{\boldmath $\sigma$})]&=
\int dx dy dx' dy'[{\hat D}(x,y){\hat D}(x',y')+{\hat E}(x,y){\hat E}(x',y')]\sum_{\mbox{\boldmath $\sigma$}} e^{s(x+y+x'+y')}\nonumber\\
&=D_0^2+E_0^2+n(2D_0D_1+2E_0E_1+S_5)+O(n^2)
\end{align}
\begin{align}
&\sum_{\mbox{\boldmath $\sigma$},\mbox{\boldmath $\tau$}}D(\mbox{\boldmath $\sigma$},\mbox{\boldmath $\tau$})E(\mbox{\boldmath $\tau$},\mbox{\boldmath $\sigma$})\prod_a[\delta(\sigma^a;-1)\delta(\tau^a;1)+\delta(\sigma^a;\tau^a)]\nonumber\\
&=\int dx dy dx' dy'{\hat D}(x,y){\hat E}(x',y') \sum_{\mbox{\boldmath $\sigma$},\mbox{\boldmath $\tau$}} e^{s(x+y')+t(y+x')}\prod_a[\delta(\sigma^a;-1)\delta(\tau^a;1)+\delta(\sigma^a;\tau^a)]\nonumber\\
&=D_0E_0+n(D_0E_1+D_1E_0+S_6)+O(n^2)
\end{align}
where $S_4=\int dx dy [{\hat D}_0(x,y)+{\hat E}_0(x,y)]\ln[2\cosh (x+y+g)]$,
$S_5=\int dx dy dx' dy'[{\hat D}_0(x,y){\hat D}_0(x',y')+{\hat E}_0(x,y){\hat E}_0(x',y')]\ln[2\cosh (x+y+x'+y')]$ and $S_6=\int dx dy dx' dy'{\hat D}_0(x,y){\hat E}_0(x',y')\ln[e^{-x+y+x'-y'}+2\cosh (x+y+x'+y')]$.

The treatment of Eqs. (B2) is slightly more involved. Consider first the
term $T=e^{g s}\sum_{\mbox{\boldmath $\tau$}_1}E(\mbox{\boldmath $\tau$}_1,\mbox{\boldmath $\sigma$})\prod_a \big[\delta(\sigma^a;-1)
\delta(\tau^a_1;-\tau^a)+\delta(\tau_1^a;\tau^a)
\delta(\tau^a;\sigma^a)\big]$ in Eq. (B2a). Performing the sum over $\mbox{\boldmath $\tau$}_1$ gives $T=\int dx dy {\hat E}(x,y)\prod_a f_{\sigma^a \tau^a}(x,y)$ with
\be
f_{\sigma^a \tau^a}(x,y)= e^{-x\tau^a-y-g}\delta(\sigma^a;-1)+e^{(x+y+g)\sigma^a}\delta(\sigma^a;\tau^a)\ .
\ee
Following Ref.\cite{BS2001}, we now use the identity
\begin{align}
\prod_a f_{\sigma^a \tau^a}=\exp[\sum_a\ln f_{\sigma^a \tau^a}]=\exp[\sum_a\sum_{\sigma,\tau}\ \delta(\sigma;\sigma^a)\delta(\tau;\tau^a) \ln f_{\sigma \tau}]
\end{align}
with
\[\sum_a\delta(\sigma;\sigma^a)\delta(\tau;\tau^a)=\left\{\begin{array}{ll} \frac{1}{4}(s+u+t+n) & \mbox{for $\sigma=\tau=1$}\\
\frac{1}{4}(s-u-t+n) & \mbox{for $\sigma=1,\tau=-1$}\\
\frac{1}{4}(-s-u+t+n) & \mbox{for $\sigma=-1,\tau=1$}\\
\frac{1}{4}(-s+u-t+n) & \mbox{for $\sigma=\tau=-1$}\ .
\end{array}
\right. \]
Remembering that the combination $\sigma^a=1, \tau^a=-1$ is forbidden and that $u=n+s-t$, we then obtain
\begin{align}
\prod_af_{\sigma^a \tau^a}(x,y)=\exp[\frac{s+n}{2} \ln f_{\ss ++}(x,y)+\frac{t-s}{2} \ln f_{\ss -+}(x,y)-\frac{t-n}{2} \ln f_{\ss --}(x,y)]
\end{align}
with $f_{\ss ++}(x,y)=e^{x+y+g}$, $f_{\ss -+}(x,y)=e^{-(x+y+g)}$ and $f_{\ss --}(x,y)=e^{x-y-g}+e^{-(x+y+g)}$. Hence
\begin{align}
T=\int dx dy {\hat E}(x,y)\exp[s(x+y+g)-\frac{t}{2}\ln(1+e^{2x})+\frac{n}{2}\ln(1+e^{2x})] \ .
\end{align}
Using the same procedure, we get
\be
e^{gs}D(\mbox{\boldmath $\sigma$},\mbox{\boldmath $\sigma$})=\int dx dy {\hat D}(x,y)\exp[s(x+y+g)] \ .
\ee
Finally, Fourier transforming both sides of Eq. (B2a) with respect to $s$ and $t$, we obtain
\begin{align}
\frac{e^{\Lambda}}{\gamma}{\hat D}(x,y)&=\frac{a-b}{2}\big\{B\delta(x-2g)\delta(y)+\delta(y)\int dx'{\hat D}(x',x-x'-g)\nonumber\\
&+\int dx'dy'{\hat E}(x',y')\delta(x-x'-y'-g)\delta(y+\frac{1}{2}\ln [1+e^{2x'}])e^{\frac{n}{2}\ln(1+e^{2x'})}\big\}
\end{align}
In the same way, the Fourier transform of Eq. (B2b) gives
\begin{align}
\frac{e^{\Lambda}}{\gamma}{\hat E}(x,y)&=\frac{b+c}{2}\big\{B\delta(x-2g)\delta(y)+\delta(y)\int dx'{\hat E}(x',x-x'-g)\nonumber\\
&+\int dx'dy'{\hat D}(x',y')\delta(x-x'-y'-g)\delta(y-\frac{1}{2}\ln [1+e^{-2x'}])e^{\frac{n}{2}\ln(1-e^{2x'})}\big\}
\end{align}

Collecting all terms at order $n^0$, we obtain the six coupled equations
\begin{subequations}
\begin{align}
\frac{e^{\Lambda_0}}{\gamma}A_0=\frac{1-a}{2}[A_0+B_0+C_0+D_0+E_0]+\frac{a-b}{2}A_0
\end{align}
\begin{align}
\frac{e^{\Lambda_0}}{\gamma}C_0=\frac{1-c}{2}[A_0+B_0+C_0+D_0+E_0]+\frac{b+c}{2}C_0
\end{align}
\be
\frac{e^{\Lambda_0}}{\gamma}B_0=\frac{a-b}{2}C_0+\frac{b+c}{2}A_0
\ee
\end{subequations}
\begin{subequations}
\begin{align}
{\hat D}_0(x,y)&=\frac{a-b}{2}\big\{B_0\delta(x-2g)\delta(y)+\delta(y)\int dx'{\hat D}_0(x',x-x'-g)\nonumber\\
&+\int dx'dy'{\hat E}_0(x',y')\delta(x-x'-y'-g)\delta(y+\frac{1}{2}\ln [1+e^{2x'}])\big\}
\end{align}
\begin{align}
{\hat E}_0(x,y)&=\frac{b+c}{2}\big\{B_0\delta(x-2g)\delta(y)+\delta(y)\int dx'{\hat E}_0(x',x-x'-g)\nonumber\\
&+\int dx'dy'{\hat D}_0(x',y')\delta(x-x'-y'-g)\delta(y-\frac{1}{2}\ln [1+e^{-2x'}])\big\}
\end{align}
\end{subequations}
and
\begin{align}
 (A_0+B_0+C_0+D_0+E_0)^2=1 \ .
\end{align}
Eqs. (B16) and (B18) may be decoupled from Eqs. (B17) by performing the integration of (B17) over $x$ and $y$ which readily gives
\be
D_0=\frac{a-b}{2-(a+c)}B_0 \ , \ E_0=\frac{b+c}{2-(a+c)}B_0 \ .
\ee
The only physical solution of Eqs. (B16), (B18) and (B19) is then
\begin{subequations}
\be
\Lambda_0=\ln \gamma
\ee
\be
A_0=\frac{1-a}{2-(a-b)}
\ee
\be
B_0=\frac{(a-b)(1-c)}{4-2(b+c)}+\frac{(b+c)(1-a)}{4-2(a-b)}
\ee
\be
C_0=\frac{1-c}{2-(b+c)} \ .
\ee
\end{subequations}
Note that Eq. (B20) is necessary for the $n\to 0$ limit to exist when $z=2$ (as pointed out after Eq. (30)). This solution corresponds to the choice $A_0+B_0+C_0+D_0+E_0=1$ in Eq. (B18) which implies  that $\lim_{n\to 0}\sum_{\mbox{\boldmath $\sigma$},\mbox{\boldmath $\tau$}}c(\mbox{\boldmath $\sigma$},\mbox{\boldmath $\tau$})=\sqrt{2/\gamma}$, a result that could have been also obtained from Eq. (21) by naively (but incorrectly) using a Fourier integral representation of $c(\mbox{\boldmath $\sigma$},\mbox{\boldmath $\tau$})$.

It remains to solve  the coupled integral equations (B17). This can be done by writting formally  ${\hat D}_0(x,y)$ and ${\hat E}_0(x,y)$ as two infinite series of delta functions. The locations of the peaks are then obtained recursively as discussed in Appendix C.   

We now achieve the calculation of the quenched complexity by solving Eqs. (B1-B3) at first order in $n$.
Using Eqs. (B4-B8), (B14-B15), (B17-B18), and (B20a), we obtain, after some manipulations,
\begin{subequations}
\begin{align}
[\Lambda_1+\frac{1-b}{2}g]A_0+\frac{1+b}{2}A_1=\frac{1-a}{2}\big[&-g(C_0+2B_0+D_0+E_0)+C_1+B_1+D_1+E1\nonumber\\
&-S1-S2+S'_3-S'_2\big]
\end{align}
\begin{align}
[\Lambda_1-\frac{1-b}{2}g]C_0+\frac{1-b}{2}C_1=\frac{1-c}{2}\big[&g(A_0+2B_0+D_0+E_0)+A_1+B_1+D_1+E_1\nonumber\\
&+S'_1+S'_2+S_3+S_2\big]
\end{align}
\be
\Lambda_1B_0+B_1=\frac{a-b}{2}C_1+\frac{b+c}{2}A_1
\ee
\end{subequations}
\begin{subequations}
\begin{align}
{\hat D}_1(x,y)&=-(y+\Lambda_1){\hat D}_0(x,y)+\frac{a-b}{2}\big\{B_1\delta(x-2g)\delta(y)+\delta(y)\int dx'{\hat D}_1(x',x-x'-g)\nonumber\\
&+\int dx'dy'{\hat E}_1(x',y')\delta(x-x'-y'-g)\delta(y+\frac{1}{2}\ln [1+e^{2x'}])\big\}
\end{align}
\begin{align}
{\hat E}_1(x,y)&=(y-\Lambda_1){\hat E}_0(x,y)+\frac{b+c}{2}\big\{B_1\delta(x-2g)\delta(y)+\delta(y)\int dx'{\hat E}_1(x',x-x'-g)\nonumber\\
&+\int dx'dy'{\hat D}_1(x',y')\delta(x-x'-y'-g)\delta(y-\frac{1}{2}\ln [1+e^{-2x'}])\big\}
\end{align}
\end{subequations}
\begin{align}
&A_1+B_1+C_1+D_1+E_1+g(C_0-A_0)B_0+\frac{B_0^2}{2}\ln[2\cosh(2g)]+(C_0-A_0)(S_2+S'_2)\nonumber\\
&+C_0(S'_1+S_3)+A_0(S'_3-S_1)+B_0S_4+\frac{S_5}{2}+S_6=0 \ .
\end{align}
Integrating Eqs. (B22) over $x$ and $y$ we obtain
\begin{subequations}
\be
D_1=\frac{a-b}{2-a-c}(B_1-\Lambda_1C_0+S'_2)-\frac{2-b-c}{2-a-c}(S_2+\Lambda_1D_0)
\ee
\be
E_1=\frac{b+c}{2-a-c}(B_1-\Lambda_1D_0-S_2)-\frac{2-a+b}{2-a-c}(S'_2-\Lambda_1D_0)
\ee
\end{subequations}
which again allows to decouple Eqs. (B21) and (B23) from (B22). The solution of the system of linear equations (B21), (B24), and (B23) then yields
\begin{align}
\Lambda_1(g)&=\frac{1}{2}\Big\{g\big [(b-c)A_0+(a+b)C_0+(a-c)(2B_0+D_0+E_0)\big ]-S_1(1-a)+S'_1(1-c)\nonumber\\
&+S_2(a-c-2)+S'_2(a-c+2)+S_3(1-c)+S'_3(1-a)\Big\}
\end{align}
which, remarkably, does not depend on the integrals $S_4, S_5$ and $S_6$. In fact, it turns out that $\Lambda_1(g)$ can be written in terms of a single integral of ${\hat D}_0(x,y)$ or ${\hat E}_0(x,y)$ (for instance $S_3$). Indeed, the analysis of Eqs. (B22) shows that

\begin{subequations}
\be
S_1=\frac{p(2-p-q)}{(1-p-q)(1-p-q+pq)}g\ , \ \ S'_1=\frac{q}{p}S_1
\ee
\be
S_2=-\frac{q}{2}(S_1+S_3) \ , \ \ S'_2=\frac{q}{2}(S_3-S_1)
\ee
\be
S'_3=\frac{q}{p}S_3 \ .
\ee
\end{subequations}
with $p=(a-b)/2$ and $q=(b+c)/2$. Inserting (B26) in Eq. (B25), using Eqs. (B19) and (B20) and the relation  $\Sigma_Q(g,H)=\Lambda_1(g)-g\partial\Lambda_1(g)/ \partial g$, we finally obtain Eq. (38).

Note that the magnetisation $m(g)$ in the limit $n\rightarrow 0$ only depends on the solution  of the saddle-point equations at lowest order. Inserting Eq. (35) in Eq. (22) gives after some algebra
\begin{align}
m(g)&= C_0^2-A_0^2 +2B_0(C_0-A_0)+2(C_0E_0-A_0D_0)+B_0^2\tanh 2g\nonumber\\
&+A_0\int dx dy {\hat E}_0(x,y)[\tanh x -1]+C_0\int dx dy {\hat D}_0(x,y)[\tanh x +1]\nonumber\\
&+2B_0\int dx dy [{\hat D}_0(x,y)+{\hat E}_0(x,y)]\tanh (x +y+g)\nonumber\\
&+\int dx dy dx' dy' [{\hat D}_0(x,y){\hat D}_0(x',y')+{\hat E}_0(x,y){\hat E}_0(x',y')]\tanh (x +y+x'+y')\nonumber\\
&+2\int dx dy dx' dy' {\hat D}_0(x,y){\hat E}_0(x',y')\frac{e^{x+y+x'+y'}-e^{-(x+y+x'+y')}}{e^{-x+y+x'-y'}+e^{x+y+x'+y'}+e^{-(x+y+x'+y'}}
\end{align}

\section{}

The solution of the coupled integral equations (B17) cannot be obtained in the form of closed expressions for ${\hat D}_0(x,y)$ and ${\hat E}_0(x,y)$.  However, this type of self-consistent equations is common in one-dimensional disordered systems and it is easy to see that  ${\hat D}_0(x,y)$ and ${\hat E}_0(x,y)$ may be represented as two {\it infinite} series of delta functions which  are conveniently written as
\begin{subequations}
\be
{\hat D}_0(x,y)/B_0=p\sum_{i=0}^{\infty}\sum_{j=0}^{i}p^jq^{i-j} {\tilde D}_{ij}(x,y)
\ee
\be
{\hat E}_0(x,y)/B_0=q\sum_{i=0}^{\infty}\sum_{j=0}^{i}q^jp^{i-j} {\tilde E}_{ij}(x,y)
\ee
\end{subequations}
with $p=(a-b)/2$, $q=(b+c)/2$ and
\begin{subequations}
\be
{\tilde D}_{ij}(x,y)=\sum_{k=1}^{\binom{i}{j}}\delta\big(x-x_k^{(ij)}(g)\big)\delta\big(y-y_k^{(ij)}(g)\big)
\ee
\be
{\tilde E}_{ij}(x,y)=\sum_{k=1}^{\binom{i}{j}}\delta\big(x-x_k'^{(ij)}(g)\big)\delta\big(y-y_k'^{(ij)}(g)\big) \ .
\ee
\end{subequations}
${\tilde D}_{ij}(x,y)$ and ${\tilde E}_{ij}(x,y)$ are then computed iteratively, the coordinates $(x_k^{(ij)},y_k^{(ij)})$ and $(x_k'^{(ij)},y_k'^{(ij)})$ of the delta peaks at step $i$ being obtained from the coordinates at step $i-1$, with ${\tilde D}_{00}(x,y)={\tilde E}_{00}(x,y)=\delta(x-2g)\delta(y)$.
For instance, the first terms in the expansion of ${\hat D}_0(x,y)$ are
\begin{align}
{\tilde D}_{10}(x,y)&=\delta(x-3g)\delta\big(y+\frac{1}{2}\ln(1+e^{4g})\big)\nonumber\\
{\tilde D}_{11}(x,y)&=\delta(x-3g)\delta(y)\nonumber\\
{\tilde D}_{20}(x,y)&=\delta(x-4g)\delta\big(y+\frac{1}{2}\ln(1+e^{6g})\big)\nonumber\\
{\tilde D}_{21}(x,y)&=\delta\big(x-4g+\frac{1}{2}\ln(1+e^{4g})\big)\delta(y)+\delta\big(x-4g-\frac{1}{2}\ln(1+e^{-4g})\big)\delta\big(y+\frac{1}{2}\ln(1+e^{6g})\big)\nonumber\\
{\tilde D}_{22}(x,y)&=\delta(x-4g)\delta(y)\nonumber\\
\mbox{etc...}
\end{align}
Note that $p$ and $q$ depend on $H$ and $R$ whereas the dependence on $g$ is only contained in the position of the delta functions. 
With Eq. (B1a), the integral $S_3$ also becomes an infinite series, 
\be
S_3(g)=B_0 p\sum_{i=0}^{\infty}\sum_{j=0}^{i}p^jq^{i-j}  \sum_{k=1}^{\binom{i}{j}}\ln[2 \cosh  x_k^{(ij)}(g)]
\ee
The number of delta functions grows exponentially during the iteration procedure and the precision on $S_3$ depends on the number $I$ of terms that are included in the infinite sums. However, the cumulative weight of the delta peaks is finite ($D_0/B_0=p/(1-p-q)$ and $E_0/B_0=q/(1-p-q)$), and one can take as many terms as needed so to satisfy a given accuracy defined by  $\epsilon=1-(1-p-q)\sum_{i=1}^I\sum_{j=0}^{i}\binom{i}{j}p^jq^{i-j}$ (see, e.g., Ref.\cite{SCS2003} for a similar calculation). 
When  $H=0$, however, one has $p=q=(1/2) \ \mbox{erf}(\sqrt{2}J/R)$ and  the convergence of the series becomes progressively poorer as the ratio $J/R$ increases and $p,q \rightarrow 1/2$. In this case, it is convenient to write $S_3(g)$ as
\be
S_3(g)=B_0 \frac{p}{1-2p}s_3^{\infty}(g)+B_0p\sum_{i=0}^{\infty}p^i[s_3^{(i)}(g)-s_3^{\infty}(g)]
\ee  
where $s_3^{(i)}(g)=\sum_{j=0}^{i}\sum_{k=1}^{\binom{i}{j}} \ln[2 \cosh  x_k^{(ij)}(g)]$ and $s_3^{\infty}(g)=\lim_{i \rightarrow \infty}s_3^{(i)}(g)$. For $g=0$, for instance, one finds numerically $s_3^{\infty}(0)\approx 0.716$
and since $B_0 \sim 1-2p$ when $R \rightarrow 0$, this yields from Eq. (38) $\lim_{R\to 0}\Sigma_Q^{max}(H=0)=s_3^{\infty}(0)/4\approx0.179$. 

When $g\rightarrow \pm \infty$, the expressions of ${\hat D}_0(x,y)$ and ${\hat E}_0(x,y)$ simplify considerably as the positions of the delta peaks concentrate on 
a few lines in the $x-y$ plane. Specifically, for $g\rightarrow +\infty$, one obtains 
\begin{subequations}
\be
{\hat D}_0(x,y)/B_0=\sum_{n=2}^{\infty} [p^{n-1}\frac{(1-p)(1-q)}{1-p-q}\delta(x-ng)\delta(y)+p{\tilde E}(n-1)\delta(x-ng)\delta(y-(1-n)g)]
\ee
\be
{\hat E}_0(x,y)/B_0=\sum_{n=2}^{\infty} {\tilde E}(n)\delta(x-ng)\delta(y)
\ee
\end{subequations}
with 
\be
{\tilde E}(n)=q\frac{(1-p)(1-q)}{1-p-q}[q^{n-2}+p\frac{p^{n-2}-q^{n-2}}{p-q}]
\ee
Replacing in Eq. (B27) gives, after some algebra,
\begin{align}
m(g\rightarrow + \infty)&=1-2A_0(1+B_0+D_0)\nonumber\\
&=1-2p(-H)
\end{align}
with 
\begin{align}
p(-H)&=\frac{(a-1)(a+ac-4+b^2-2b-c)}{(2-a+b)^2}\nonumber\\
&=p_0\frac{1+p_0p_2-p_1^2}{[1+p_0-p_1]^2}
\end{align}
where $p_n(-H)=\int_{2(1-n)+H}^{\infty}P(h)dh$  is the probability (introduced in Ref.\cite{S1996}) that a spin with $n$ neighbors up  ($n=0,1,2$) is up at the field $-H$: $p_0=(1-a)/2,p_1=(1-b)/2,p_2=(1+c)/2$. The magnetization given by Eqs. (C8-C9) is exactly that calculated in Ref.\cite{S1996} for the upper branch of the hysteresis loop. By symmetry, the magnetization on the lower branch of the loop is given by $m=2p(H)-1$ and corresponds to the limit $g\rightarrow -\infty$.

It remains to calculate $\Lambda_1(g,H)$ and $\Sigma_Q(g,H)=\Lambda_1(g,H)-gm=\Lambda_1(g,H)-g\partial \Lambda_1(g,H)/\partial g$ when $g\rightarrow \pm \infty$. Inserting Eqs. (C6-C7) in (B25) gives, after some algebra, 
\be
\Lambda_1(g\rightarrow + \infty,H)\sim -g \frac{(a^2c-a-2ac+ab^2-2b^2-2b+c)}{(2-a+b)^2} \ , 
\ee
from which one readily finds that $\Sigma_Q(g \rightarrow + \infty,H)=0$. By symmetry,
$\Sigma_Q$ also vanishes along the lower branch of the hysteresis loop.
To illustrate how the cancellation of terms occurs in $\Sigma_Q(g,H)$ in the limit $g\rightarrow \pm \infty$, let us give the first terms in the large-$R$ expansion of $\Lambda_1(g,H)$ and $\Sigma_Q(g,H)$:

\begin{align}
\Lambda_1(g,H)&=g\sqrt{\frac{2}{\pi}} \frac{H}{R}+ \frac{2J}{\pi R^2}[2gH+J\ln(2\cosh 2g)]-\frac{1}{6\pi R^3}\sqrt{\frac{2}{\pi}}[\pi g H^3+6\pi g H J^2-24g H J^2\nonumber\\
&+24J^3 \ln (2\cosh 2g) -24 J^3 \ln (2 \cosh 3g)]  +O(\frac{J^4}{R^4})
\end{align}

\begin{align}
\Sigma_Q(g,H)&=\frac{2J^2}{\pi R^2}[\ln(2\cosh2g)-2g \tanh 2g]+\frac{4 J^3}{\pi R^3}\sqrt{\frac{2}{\pi}}[2g\tanh 2g-\ln (2\cosh 2g)\nonumber\\
& +\ln (2 \cosh 3g) -3g \tanh 3g] +O(\frac{J^4}{R^4})
\end{align}
Note that $\Sigma_Q(g,H)$ does not depend on $H$ at the first two orders of the expansion.


\begin{thebibliography}{10}
\bibitem{S1993} J. P. Sethna, K. Dahmen, S. Kartha, J. A. Krumhansl,
B. W. Roberts, and J. D. Shore, Phys. Rev. Lett. {\bf 70}, 3347
(1993). 
\bibitem{DS1996} K. Dahmen and J. P. Sethna, Phys. Rev. B {\bf  53},
14872 (1996); O. Perkovic, K. Dahnen, and J. P. Sethna, Phys. Rev. B
{\bf 59}, 6106 (1999).
\bibitem{S1996} P. Shukla,  Physica A {\bf  233}, 235  (1996).
\bibitem{DSS1997} D. Dhar, P. Shukla, and J. P. Sethna, J. Phys. A
{\bf  30}, 5259 (1997).
\bibitem{BIJPB2000} A. Berger, A. Inomata, J. S. Jiang, J. E. Pearson, and S. D. Bader, Phys. Rev. Lett. {\bf 85}, 4176
(2000).
\bibitem{M2003} J. Marcos, E. Vives, Ll. Ma\~{n}osa, M. Acet, E. Duman,
M. Morin, V. Novak, and A. Planes, Phys. Rev. B {\bf 67}, 224406
(2003).
\bibitem{DKRT2003} F. Detcheverry, E. Kierlik, M. L. Rosinberg, and
G. Tarjus, Phys. Rev. E {\bf 68}, 061504 (2003); Langmuir {\bf 20},  8006  (2004).
\bibitem{TYC1999} D. J. Tulimieri, J. Yoon, and M. H. W. Chan, Phys. Rev. Lett. {\bf 82}, 121 (1999).
\bibitem{note00} In this work, we study the $1$-spin-flip stable states whose energy cannot be lowered by the flip of any single spin. The corresponding dynamics consists in aligning each spin with its local field.  Generalization to $2$-spin-flip stable states and associated dynamics is considered in E. Vives, M.L. Rosinberg, G. Tarjus, 
preprint cond-mat/0411330 (2004).
\bibitem{PZZ1999} F. Pazmandi, G. Zarand, and G. Zimanyi,
Phys. Rev. Lett. {\bf 83}, 1034 (1999).
\bibitem{L2000} A. A. Likhachev, preprint cond-mat/0007504 (2000).
\bibitem{CGZ2002} F. Colaiori, A. Gabrielli, and S. Zapperi,
Phys. Rev. B {\bf 65}, 224404 (2002). 
\bibitem{MP2001} M. Mezard and G. Parisi, Eur. Phys. J. {\bf B 20},
217 (2001).
\bibitem{LD2001} A. Lef\`evre and D. S. Dean, Eur. Phys. J. {\bf B 21}, 121 (2001).
\bibitem{BS2001} J. Berg and M. Sellito, Phys. Rev. E {\bf 65}, 016115
(2001).
\bibitem{PPR2002} A. Pagnani, G. Parisi, and M. Rati\'eville,
Phys. Rev. E {\bf 67}, 026116 (2003).
\bibitem{D2000} D. S. Dean, Eur. Phys. J. {\bf B 15}, 493 (2000).
\bibitem{MJWS1993} S. Masui, A. E. Jacobs, C. Wicentowich, and
B. W. Southern, J. Phys. A  {\bf 26}, 25 (1993).
\bibitem{note1} More precisely, we expect the complexities, as defined below, to be equal. It was checked numerically in Ref.\cite{DSS1997} that the hysteresis loops are identical in the thermodynamic limit.
\bibitem{VB1985} L. Viana and A. J. Bray, J. Phys. C. {\bf 18}, 3037 (1985).
\bibitem{BM2004} V. Basso and A. Magni, private communication.
\bibitem{SP1977} T. Schneider and E. Pytte, Phys. Rev. B {\bf 15}, 1519 (1977).
\bibitem{F2003} S. Franz, private communication.
\bibitem{note4} Preliminary numerical simulations of the $1$-d RFIM show that the typical  magnetization $m(g=0,H)$ of the metastable states is (slightly) different from the mean magnetization  of the states that are obtained dynamically by starting from random initial configurations of maximal entropy (i.e. with $s_i=\pm 1$ with probability $1/2$); this latter process corresponds to an instantaneous ``quench" of the system from an infinite temperature to $T=0$. This result (if confirmed) implies that the basins of attraction of the metastable states do not have the same size under the one-spin-flip dynamics.
\bibitem{SCS2003} R. O. Sokolovskii, M. E. Cates, and T. G. Sokolovska, Phys. Rev. E {\bf 68}, 026124 (2003).
\bibitem{BM2004b} V. Basso and A. Magni, Physica B {\bf 343}, 275 (2004).
\bibitem{note10} Note that the reverse (inner) trajectories which bound the two domains of existence of the $H$-states start from the last available states on the convex parts of the major loop (spinodals) and meet the concave, inaccessible parts of that loop at the two singular points with an infinite slope. The  fields at the starting and meeting points are separated by $2J$, which shows that the proof given by P. Shukla in Phys. Rev. E, {\bf 63}, 27102 (2001) is applicable even in the case of a discontinuous hysteresis loop.

\end{thebibliography}
\end{document}